\documentclass[12pt,reqno]{amsart}

\usepackage[OT1]{fontenc}
\usepackage{amsthm,amsmath,amsfonts,amssymb,amsaddr}
\usepackage{xr-hyper}
\usepackage{xr}
\usepackage[colorlinks=true,linkcolor=black, citecolor=blue, urlcolor=blue,bookmarks=false]{hyperref}
\usepackage{mathrsfs}
\usepackage{ascmac}
\usepackage{bm}
\usepackage{fancybox}
\usepackage{graphicx}
\usepackage{xcolor}
\usepackage{tabularx}
\usepackage{subcaption}
\usepackage[plain,noend]{algorithm2e}
\usepackage{comment}
\usepackage{fancyhdr}
\usepackage[semicolon, sort,comma,authoryear,round]{natbib}
\usepackage{fullpage}
\usepackage{newtxtext,newtxmath}
\usepackage{csvsimple}
\usepackage{multirow}

\theoremstyle{definition}

\newcommand{\Ep}{\mathbb{E}}
\renewcommand{\Pr}{\mathbb{P}}

\newcommand{\T}{\mathrm{\scriptscriptstyle T}}

\renewcommand{\hat}{\widehat}
\renewcommand{\tilde}{\widetilde}

\newcommand{\given}{\,|\,}

\DeclareMathOperator{\Var}{Var}
\DeclareMathOperator{\Cov}{Cov}

\DeclareMathOperator{\vect}{vec}

\makeatletter
\newcommand*{\addFileDependency}[1]{
\typeout{(#1)}
\@addtofilelist{#1}
\IfFileExists{#1}{}{\typeout{No file #1.}}
}\makeatother

\newcommand*{\myexternaldocument}[1]{%
\externaldocument{#1}%
\addFileDependency{#1.tex}%
\addFileDependency{#1.aux}%
}
\myexternaldocument{supplementary}

\title{Spatial Confounding in Multivariate Areal Data Analysis}
\author{Kyle Lin Wu and Sudipto Banerjee}
\date{\today}

\begin{document}

\begin{abstract}
We investigate spatial confounding in the presence of multivariate disease dependence. In the ``analysis model perspective" of spatial confounding, adding a spatially dependent random effect can lead to significant variance inflation of the posterior distribution of the fixed effects. The ``data generation perspective" views covariates as stochastic and correlated with an unobserved spatial confounder, leading to inferior statistical inference over multiple realizations. Although multiple methods have been proposed for adjusting statistical models to mitigate spatial confounding in estimating regression coefficients, the results on interactions between spatial confounding and multivariate dependence are very limited. We contribute to this domain by investigating spatial confounding from the analysis and data generation perspectives in a Bayesian coregionalized areal regression model. We derive novel results that distinguish variance inflation due to spatial confounding from inflation based on multicollinearity between predictors and provide insights into the estimation efficiency of a spatial estimator under a spatially confounded data generation model. We demonstrate favorable performance of spatial analysis compared to a non-spatial model in our simulation experiments even in the presence of spatial confounding and a misspecified spatial structure. In this regard, we align with several other authors in the defense of traditional hierarchical spatial models \citep{khanRethinkingSpatialConfounding2023, zimmermanDeconfoundingSpatialConfounding2022, gilbertConsistencyCommonSpatial2025} and extend this defense to multivariate areal models. We analyze county-level data from the US on obesity / diabetes prevalence and diabetes-related cancer mortality, comparing the results with and without spatial random effects.
\end{abstract}

\maketitle
\section{Introduction}\label{sec:introduction}

When do spatial effects distort statistical inference for fixed effects? \cite{hodgesAddingSpatiallyCorrelatedErrors2010} introduced this question in the context of a linear mixed-effects model, arguing that adding a spatially dependent random effect can induce increased estimation bias or variance inflation. They call this phenomenon ``spatial confounding.'' This notion challenges the conventional wisdom that a spatially dependent random effect should be added when residuals in a non-spatial model exhibit spatial autocorrelation. To mitigate spatial confounding, \cite{hodgesAddingSpatiallyCorrelatedErrors2010} proposed restricting the spatial effect to the orthogonal complement of the column space of the covariates, which preserves the non-spatial estimates, but faces criticism for inferior inference \citep{khanRestrictedSpatialRegression2022, zimmermanDeconfoundingSpatialConfounding2022}.

Subsequent articles on spatial confounding have typically adopted a ``data analysis'' or a ``data generation'' perspective \citep{khanRethinkingSpatialConfounding2023}. Following \cite{hodgesAddingSpatiallyCorrelatedErrors2010}, spatial confounding arises in the analysis perspective from empirical associations between observed explanatory variables and 
a spatial random effect incorporated into the model to account for spatial variability in the response unexplained by the covariates.
The proposed adjustments include restricted spatial regression \citep{hodgesAddingSpatiallyCorrelatedErrors2010, hughesDimensionReductionAlleviation2013}, restricted spatial GEE \citep{huiSpatialConfoundingGeneralized2022}, and SPOCK \cite{pratesAlleviatingSpatialConfounding2019}. In the generation perspective, spatial confounding occurs when covariates $X$ are stochastic, spatially dependent, and correlated with an unobserved spatial confounder $Z$ \citep{paciorekImportanceScaleSpatialConfounding2010}. Effects on inference are evaluated in realizations of the observed dataset. Recent methods that address spatial confounding from a data generation perspective include \texttt{Spatial+} \citep{dupontSpatialNovelApproach2022}, gSEM \citep{thadenStructuralEquationModels2018} and a spectral adjustment introduced by \cite{guanSpectralAdjustmentSpatial2023}. \cite{bolinSpatialSelfconfoundingSmoothnessrelated2025} propose covariate smoothing to address generation spatial "self-confounding" resulting from covariate misspecification.

This manuscript provides new insights into spatial confounding in the context of Bayesian multivariate spatial regression. Recently, \cite{gilbertConsistencyCommonSpatial2025} proved the consistency of the generalized least squares spatial estimator under a causal inference framework with random sampling locations and a spatial confounding criterion for data generation. We offer an alternative defense of traditional spatial estimation 
from both perspectives for finite samples and fixed sampling locations.
Furthermore, studies exploring how multivariate dependencies interact with spatial confounding are somewhat limited; \cite{urdangarinSimplifiedSpatialApproach2024} analyze the Spatial+ adjustment method in multivariate settings, 
while \cite{azevedoMSPOCKAlleviatingSpatial2021} propose MSPOCK, a multivariate extension that combines the SPOCK framework with a shared component model. We analyze multivariate dependencies in a coregionalized areal model following \cite{jinOrderFreeCoRegionalizedAreal2007} and provide theoretical results on the dependence between variables under maximal spatial smoothing. Our results align with several other articles on the defense of traditional hierarchical spatial models \citep{khanRethinkingSpatialConfounding2023, zimmermanDeconfoundingSpatialConfounding2022, gilbertConsistencyCommonSpatial2025}. We extend the results therein to multivariate areal models.

%
Subsequently, Section~\ref{sec:dg_analysis_models} defines the multivariate data generation and Bayesian analysis models. Section~\ref{sec:multivariate_confounding} examines the effect of adding a spatial random effect on the posterior distribution of fixed effects in the analysis model and estimation properties under the generation model. Section~\ref{sec:simulation} evaluates the analysis models over simulated datasets on a map of Californian counties with a spatial confounder. Section~\ref{sec:application} applies the multivariate Bayesian spatial model to estimates of obesity and diabetes prevalence rates at the county level in the US alongside diabetes-related cancer mortality rates. Section \ref{sec:discussion} concludes with a discussion.

\section{Data Generation and Analysis Models}\label{sec:dg_analysis_models}

\subsection{Notation and Definitions}\label{subsec:notation}

We follow earlier work \citep{zimmermanDeconfoundingSpatialConfounding2022, khanRethinkingSpatialConfounding2023, gilbertConsistencyCommonSpatial2025} by distinguishing between the true data generation process and the analysis model used to estimate association between $y_i$, the $n \times 1$ vector corresponding to the $i$th outcome, and the $p$ covariates, $x_{1}, \ldots, x_{p}$, for $i = 1, \ldots k$. Separation of the data generation model with fixed, unknown parameters from the Bayesian analysis model permits examination of model misspecification effects on 
the frequentist properties of the Bayesian spatial and non-spatial estimators. Table~\ref{tab:data_generation_notation} and Table~\ref{tab:analysis_notation} list the variables and parameters in the data generation and analysis models, respectively. Matrices are denoted using uppercase English and Greek letters while vectors and scalars are denoted using lowercase letters. 

\begin{table}[]
    \centering
    \begin{tabular}{cccc}
    \hline
         Variable & Dimension & Random? & Description\\
    \hline
         $Y$ & $n \times k$ & Yes & Outcome matrix \\
         $X$ & $n \times (p + 1)$ & Yes & Design matrix with intercept ($X = [1_n, X_1]$) \\
         $X_1$ & $n \times p$ & Yes & Design matrix \\
         $B$ & $(p + 1) \times k$ & No & Outcome model covariate regression coefficients \\
         $Z$ & $n \times k$ & Yes & Matrix of spatial confounders \\
         $D$ & $p \times k$ & No & Confounder model covariate regression coefficients \\
         $V_{\phi}$ & $n \times n$ & No &  Data generation model latent spatial covariance matrix \\
         $A$ & $k \times k$ & No &  Between-outcome dependence matrix factor of $E_{Y}, E_{Z}$ \\ 
         $C$ & $p \times p$ & No & Between-outcome dependence matrix factor of $E_{X}$ \\
         $\eta_{j}$ & $n \times 1$ & Yes & Outcome model $j$th latent error vector \\
         $E_{Y}$ & $n \times k$ & Yes & Outcome model latent error matrix \\
         $\zeta_j$ & $n \times 1$ & Yes & Spatial confounder model $j$th latent error vector \\
         $E_{Z}$ & $n \times k$ & Yes & Spatial confounder model latent error matrix \\
         $\epsilon_{k}$ & $n \times 1$ & Yes & Covariate model $k$th latent error vector \\
         $E_{X}$ & $n \times p$ & Yes & Covariate model latent error matrix\\
    \hline
    \end{tabular}
    \caption{Mathematical notation for variables in the data generation model. Parameters are considered as fixed and unknown quantities to be estimated in the analysis model.}
    \label{tab:data_generation_notation}
\end{table}

In the data generation model, we assume $y_i$ is associated with $z_i$, a $n \times 1$ vector of unknown confounding effects, and the $n \times p$ covariate matrix $X_1 = [x_{1}, \ldots, x_{p}]$ according to the model,
\begin{equation}\label{eqn:dg_model_single_outcome}
    \underbrace{y_{i} = \beta_{0i} 1_n + X_1 \beta_{1i} + z_{i} + \sum_{j = 1}^{k} a_{ji}\eta_{j}}_{p(y_{i} \given X_1, z_i)}, \;\;  \underbrace{z_{i} = \delta_{0i} 1_n + X_1 \delta_{1i} + \sum_{j = 1}^k a_{ji}\zeta_{i}}_{p(z_i \given X_1)}, \;\;
    \underbrace{x_{j} = \mu_j 1_n + \sum_{k = 1}^p c_{kj}\epsilon_{j}}_{p(X_{1j})},
\end{equation}
for each $i = 1, \ldots, k$ and $j = 1, \ldots, p$. Scalar intercepts for $y_i$, $z_{i}$, and $x_{j}$ are denoted by $\beta_{0i}$, $\delta_{0i}$, and $\mu_j$, respectively; $1_n$ is the $n \times 1$ vector of ones; and $\beta_{1i}$ and $\delta_{1i}$ are $p \times 1$ vectors of regression coefficients. The intercepts and coefficients are fixed, unknown quantities;
the $n \times 1$ vectors $\eta_i$, $\zeta_i$, and $\epsilon_j$ are random zero mean latent error terms while $a_{ji}$ and $c_{ji}$ are fixed constants. In Section~\ref{sec:multivariate_confounding}, we subscript expectations as $\Ep_{DG}[\cdot]$ and variances as $\Var_{DG}(\cdot)$ to indicate evaluation under the data generation model.

\begin{table}[]
    \centering
    \begin{tabular}{ccc}
    \hline
         Variable & Dimension & Description\\
    \hline
         $Y$ & $n \times k$ & Outcome matrix \\
         $X$ & $n \times (p + 1)$ & Design matrix with intercept \\
         $B_S$ & $(p + 1) \times k$ & Covariate regression coefficients (spatial analysis model)\\
         $G$ & $n \times k$ & Spatial random effects \\
         $\Phi$ & $n \times k$ & Latent spatial factors \\
         $\Psi$ & $n \times k$ & Latent non-spatial factors \\
         $M$ & $k \times k$ & Between-outcome matrix factor (spatial analysis model) \\
         $R$ & $k \times k$ & Diagonal matrix of spatial variance proportions in latent factors \\
         $W_{\phi}$ & $n \times n$ & Analysis model latent spatial covariance matrix \\
         $E_{S}$ & $n \times k$ & Residual error matrix (spatial analysis model) \\
         $B_{NS}$ & $(p + 1) \times k$ & Covariate regression coefficients (non-spatial analysis model) \\
         $E_{NS}$ & $n \times k$ & Residual error matrix (non-spatial analysis model) \\
         $\Sigma_{NS}$ & $n \times k$ & Between-outcome covariance matrix (non-spatial analysis model)\\
    \hline
    \end{tabular}
    \caption{Mathematical notation for variables in Bayesian non-spatial and spatial analysis models.}
    \label{tab:analysis_notation}
\end{table}

In Bayesian analysis, the confounders $z_1, \ldots, z_k$ are not observed and hence the direct effect of the covariates $X_1$ on $y_i$ is obscured. Since the association between $X_1$ and $y_i$ is of primary interest, only a model for $[y_1, \ldots, y_k \given X_1]$ is specified.

\subsection{Data Generation Model with Spatial Confounder}\label{subsec:dg_model}

For the data generation model in \eqref{eqn:dg_model_single_outcome}, we focus on spatially dependent $X_1$ and $Z$ by introducing spatial structure in $\Var(\zeta_{i})$ and $\Var(\epsilon_i)$. For a univariate outcome, the BYM (Besag, York and Mollié) model and the BYM2 reparameterization introduced in \cite{rieblerIntuitiveBayesianSpatial2016a} are flexible classes of spatial analysis models that incorporate spatial and non-spatial error variances. Following \cite{jinOrderFreeCoRegionalizedAreal2007}, we extend the BYM2 model to the multivariate setting by modeling the error terms of $z_{i}$ and $x_{j}$ as linear transformations of independent latent univariate BYM2 spatial processes. Let $V_\phi$ be a fixed positive definite $n \times n$ matrix that encodes spatial autocorrelation, $\rho_i \in [0, 1]$, and $0_n$ be an $n \times 1$ vector of zeroes. We assume the error terms $\eta_i, \zeta_i, \epsilon_i$ are generated independently as
\begin{equation}\label{eqn:dg_error_variance}
    \eta_i \sim N_{n}(0_n, (1 - \rho_i) I_n), \quad\zeta_i \sim N_{n}(0_n, \rho_i V_{\phi}), \quad\epsilon_{j} \sim N_{n}(0_n, V_{\phi}),
\end{equation}
for $i = 1, \ldots, k$ and $j = 1, \ldots, p$. Since variation in $y_i$ conditional on $X_1$ stems from $\sum_{j = 1}^k a_{ij}(\epsilon_i + \zeta_i)$ and $\Var(\epsilon_i + \zeta_i) = (1 - \rho_i)I_n + \rho_i V_{\phi}$, $\rho_i$ can be interpreted as the spatial proportion of the error variance in the $i$th latent spatial process conditional on $X_1$ when the geometric mean of the marginal prior variances in $V_\phi$ is one \citep{rieblerIntuitiveBayesianSpatial2016a}.

Letting $Y = [y_1, \ldots, y_k]$ and $Z = [z_1, \ldots, z_k]$ be $n\times k$ matrices, we cast $\eqref{eqn:dg_model_single_outcome}$ as
\begin{equation}\label{eqn:data_generating_model}
    Y = 1_n \beta_0^{\T} + X_1B_1 + Z + E_{Y} A, \quad
    Z = 1_n \delta_0^{\T} + X_1 D_1 + E_{Z} A,  \quad
    X_1 = 1_n \mu^{\T} + E_{X} C,
\end{equation}
where $B_1$ and $D_1$ are $p \times k$ with $i$th columns $\beta_{1i}$ and $\delta_{1i}$, respectively; $\beta_0 = (\beta_{01}, \ldots, \beta_{0k})^{\T}$ and $\delta_0 = (\delta_{01}, \ldots, \delta_{0k})^{\T}$ are $k\times 1$; and $A = (a_{ji})$ and $C = (c_{kj})$ are $k \times k$ and $p \times p$, respectively, and assumed to be invertible. $A$ and $C$ function as both scaling and linear transformations from the latent error terms to the observed data.  The random residual error matrices $E_Y = [\eta_1, \ldots, \eta_{k}]$ and $E_Z = [\zeta_1, \ldots, \zeta_{k}]$ are $n \times k$ while $E_{X} = [\epsilon_{1}, \ldots, \epsilon_{p}]$ is $p \times p$. Since $\Cov(z_i, \vect(X_1)) = (\delta_{1i}^{\T}C^{\T}C) \otimes V_{\phi}$ where $\otimes$ denotes the Kronecker product and $\vect(X_1)$ is obtained by stacking the columns of $X_1$, we define $z_i$ as a spatial confounder if $\delta_{1i} \neq 0_p$. 

\subsection{Spatial and Non-Spatial Bayesian Analysis Models}\label{subsec:analysis_models}

\subsubsection{Non-Spatial Bayesian Analysis}\label{subsubsec:nonspatial_analysis}

We estimate $B = (\beta_0, B_1^{\T})^{\T}$ using a non-spatial conjugate Bayesian multivariate regression model,
\begin{equation}\label{eqn:nonspatial_analysis_model}
    Y = XB_{NS} + E_{NS}, \;\; \vect(E_{NS}) \sim \mbox{N}_{nk}(0_{nk}, \Sigma_{NS} \otimes I_n),\;\; \pi(B_{NS}, \Sigma_{NS}) \propto \mbox{IW}(\Sigma_{NS} \given v\Sigma_0, v),
\end{equation}

where the design matrix $X = (1_n, X_1)$ is full column rank, $B_{NS}$ is the $(p + 1) \times k$ matrix of intercepts and slopes, $\Sigma_{NS}$ is $k \times k$ and encodes between-outcome covariance, and the prior $\pi(B_{NS}, \Sigma_{NS})$ is uninformative on $\beta$ while placing an inverse-Wishart distribution on $\Sigma_{NS}$ with degrees of freedom $v > k - 1$ and $k \times k$ scale matrix $\Sigma_0$ specifying the mean $\Ep[\Sigma_{NS}^{-1}] = \Sigma_{0}^{-1}$. When $\Sigma_{NS} = I_k$, the outcomes $y_1, \ldots, y_k$ are modeled as independent. We denote parameters in the analysis model with a subscript to emphasize a separation from the fixed unknown parameters in the data generating model in \eqref{eqn:data_generating_model}. The joint posterior is
\begin{equation}
    \label{eq: normal_wishart_posterior}
    \pi(B_{NS}, \Sigma_{NS} \given Y, X_1) = \underbrace{\mbox{IW}(v_\star, \Sigma_\star)}_{p(\Sigma_{NS}\given Y, X_1)}\times \underbrace{\mbox{N}_{(p + 1)k}\left(\vect\left(\hat{B}_{NS}\right), \Sigma_{NS} \otimes (X^{\T}X)^{-1}\right)}_{p(\vect(B_{NS})\given \Sigma_{NS}, Y, X_1)}\;,
\end{equation}

where $\hat{B}_{NS} = (X^{\T}X)^{-1}X^{\T}Y$, $v_\star = v + n$ and $\Sigma_{\star} = v\Sigma_0 + (Y - X\hat{B}_{NS})^{\T}(Y - X\hat{B}_{NS})$. Exact samples from $\pi(B_{NS}, \Sigma_{NS} \given Y, X_1)$ are obtained by simulating $\Sigma_{NS} \sim \mbox{IW}(v_\star, \Sigma_\star)$ and then drawing 
$\vect(B_{NS}) \sim \mbox{N}_{(p + 1)k}\left(\vect\left(\hat{B}_{NS}\right), \Sigma_{NS} \otimes (X^{\T}X)^{-1}\right)$ for each 
value of $\Sigma_{NS}$.


\subsubsection{Spatial Bayesian Analysis}

Turning to spatial analysis, we consider a multivariate BYM2 parameterization of a spatial random effect by writing the spatial analysis model as
\begin{equation}\label{eqn:spatial_analysis_model}
\begin{split}
    Y &= XB_{S} + G + E_S, \quad G = \Phi M, \quad E_S = \Psi M, \quad \Phi = (\phi_1, \ldots, \phi_k), \quad \Psi = (\psi_1, \ldots, \psi_k),\\ 
    \phi_{i} &\stackrel{\text{ind.}}{\sim} N_{n}(0_n, r_{i} W_{\phi}), \quad\psi_{i}  \stackrel{\text{ind.}}{\sim} N_{n}(0_n, (1 - r_{i})I_n) \;\text{ for $i = 1, \ldots, k$,}  
\end{split}
\end{equation} 
where the regression coefficient matrix $B_{S}$ is $(p + 1) \times k$, $G$ is $n \times k$ comprising spatial effects, $M$ is $k \times k$ and invertible that induces correlation between outcomes, $W_\phi$ is known $n \times n$ and positive definite capturing spatial autocorrelation, and $R = \mbox{diag}(r_1, \ldots, r_k)$ with $r_i \in [0, 1]$ for $i = 1, \ldots k$. 
The spatial residuals $G$ and the non-spatial residuals $E_S$ are modeled in the prior as linear transformations through $M$ of the spatial latent factors $\phi_1, \ldots, \phi_k$ and non-spatial latent factors $\psi_1, \ldots, \psi_k$, respectively, allowing for both spatial and non-spatial heterogeneity. Fixing $M = I_k$ results in modeling the outcomes $y_1, \ldots, y_k$ as independent. Similar to the role of $\rho_i$ in \eqref{eqn:data_generating_model}, $r_i$ represents the spatial proportion of variance in the $i$th latent process, and setting $R = O$ reduces to the non-spatial likelihood in \eqref{eqn:nonspatial_analysis_model}. $W_{\phi}^{-1}$ represents the assumed spatial structure; common constructions include the conditionally autoregressive model (CAR), simultaneous autoregressive model (SAR), or the directed acyclic graph auto-regressive model (DAGAR) developed in \cite{dattaSpatialDiseaseMapping2019b}. Despite accounting for spatial heterogeneity in the data generation process, \eqref{eqn:spatial_analysis_model} is misspecified because the prior on $G$ does not account for the dependency between the confounders $Z$ and covariates $X$.


We specify the prior for \eqref{eqn:spatial_analysis_model} as $\pi(B_{S}, G, M, R) = N_{nk}(\vect(G) \given 0_{nk}, M^{\T}RM \otimes W_{\phi}) \times \pi(M)\times\pi(R)$. For $\pi(M)$, we let $\Sigma_{S} = M^{\T} M$, where $M$ is the upper-triangular Cholesky factor, and set $\pi(\Sigma_S) = \mbox{IW}(\Sigma_S \given v\Sigma_0, v)$ 
such that $\Ep[\Sigma^{-1}_S]  = \Sigma^{-1}_0$. Hence, $\pi(M) \propto \mbox{IW}(M^{\T}M \given v\Sigma_0, v) \times \Pi_{i = 1}^k M_{ii}^{k - i + 1}$, where $\Pi_{i = 1}^k M_{ii}^{k - i + 1}$ is proportional to the Jacobian $\left\vert \frac{\partial \Sigma_S}{\partial M_{ij}}\right\vert$. Following \cite{simpsonPenalisingModelComponent2017a}, we specify a penalized complexity (PC) prior $\pi(R) = \mbox{PC}(R \given \lambda_R) \propto \lambda_R \exp(-\lambda_R \sqrt{2\text{KLD}(R)})$ that discourages oversmoothing, where $\lambda_R > 0$ is a flexibility hyperparameter and $\text{KLD}(R)$ is the Kullback-Leiber divergence of $\mbox{N}_{nk}(0, (I_k - R) \otimes I_n + R \otimes W_{\phi})$ from $\mbox{N}_{nk}(0, I_{nk})$. Here, $2\text{KLD}(R) = -n \sum_{i = 1}^{k} r_i + \left(\sum_{i = 1}^k r_i \right)\left(\sum_{j = 1}^n \lambda_i^{-1}\right) - \sum_{i = 1}^k \sum_{j = 1}^n \log (1 - r_i + r_i\lambda_j^{-1})$, where $\lambda_1, \ldots, \lambda_n$ are the eigenvalues of $W_\phi^{-1}$. Small values of $\lambda_R$ represent a vague prior while larger values of $\lambda_R$ induce heavier shrinkage towards the non-spatial model. A probabilistic statement such as $\Pr(\sum_{i = 1}^{n} r_i \leq V) = q$ where $V, q \in (0, 1)$ can used as a heuristic to derive $\lambda_R$ \citep{simpsonPenalisingModelComponent2017a}. Samples can be obtained from $\mbox{PC}(R \given \lambda_R)$ via rejection sampling by repeatedly drawing $r_1^{\star}, \ldots, r_k^{\star} \overset{i.i.d.}\sim \mbox{Unif}(0, 1)$ and accepting $R_{\star} = \mbox{diag}(r_1^{\star}, \ldots, r_k^{\star})$ with probability $\exp(-\lambda_{R}\sqrt{2\mbox{KLD}(R_{\star}})$. In the simulation and data analysis sections, we use $\lambda_R = 0.01$, corresponding to a vague prior on $R$. The complete prior is
\begin{equation}\label{eqn:full_spatial_prior}
    \begin{split}
        \pi(B_S, G, M, R) &\propto \mbox{N}_{nk}(\vect(G) \given 0, M^{\T}RM \otimes W_{\phi}) \\
    &\qquad \times \mbox{IW}(M^{\T}M \given v, v\Sigma_0) \times \mbox{PC}(R \given \lambda_R) \times \Pi_{i = 1}^k M_{ii}^{k - i + 1}.
    \end{split}
\end{equation}

For tractable analysis, we contrast the posterior profile of $B_{S}$ conditioned on $M$ and $R$ to the posterior of $B_{NS}$ in the non-spatial model of \eqref{eqn:nonspatial_analysis_model} to study the effect of adding the spatial effect $G$ to the analysis model on point estimation and uncertainty quantification. 
\subsection{Estimation of Fixed Effects}\label{subsec:model_estimation}

Under the non-spatial analysis model in \eqref{eqn:nonspatial_analysis_model}, a natural point estimate for $B$ is $\hat{B}_{NS} = \Ep[B_{NS} \given Y, X_1] = (X^{\T}X)^{-1}X^{\T}Y$, which is equivalent to separate univariate ordinary least squares estimates for each outcome. For the spatial analysis model given by \eqref{eqn:spatial_analysis_model}, let $H = X(X^{\T}X)^{-1}X^{\T}$
and $\hat{G} = \Ep[G \given Y, M, R, X_1]$. Section~\ref{append:spatial_condpost_derivation} derives the conditional posterior $\pi(B_{S}, \gamma \given Y, M, R, X_1)$ and 
the conditional spatial estimator $\tilde{B}_{S} = \Ep[B_S \given Y, M, R, X_1]$ is
\begin{equation}\label{eqn:spatial_beta_estimate}  
        \tilde{B}_{S} = \hat{B}_{NS} - (X^{\T}X)^{-1}X^{\T}\hat{G}, \quad
        \vect({\hat{G}}) = (M^{\T} \otimes I_n)L^{-1} \vect\left((I_n - H)YM^{-1}\right),
\end{equation}
and $L = I_k \otimes (I_n - H) + (R^{-1}(I_k - R)) \otimes W_{\phi}^{-1}$. The term $(X^{\T}X)^{-1}X^{\T}\hat{G}$ is the adjustment to the estimation of $B$ from adding a spatial effect that is composed of an inseparable interaction of latent spatial autocorrelation and dependence between results through $W_\phi^{-1}$, $M$, and $R$. The spatial effects $G$ absorb spatial information from $Y$ through the residuals $(I_n - H)Y$, as $\tilde{B}_{S}$ can be interpreted as the estimate of $B$ under the non-spatial model in \eqref{eqn:nonspatial_analysis_model} with $Y_{\star} = Y - \Ep[G \given Y, M, R, X_1]$ as the response. 
The non-spatial estimator can be recovered under the spatial analysis model in \eqref{eqn:spatial_analysis_model} by the relation $\hat{B}_{NS} = \Ep[B_{S} + (X^{\T}X)^{-1}X^{\T}G \given Y, M, R, X_1]$. 

\section{Multivariate Spatial Confounding}\label{sec:multivariate_confounding}

\subsection{Analysis Model Posterior Variance Inflation}\label{subsec:analysis_confounding}

We first compare posterior inference under the Bayesian spatial analysis model of \eqref{eqn:spatial_analysis_model} and non-spatial analysis model of \eqref{eqn:nonspatial_analysis_model}. \cite{hodgesAddingSpatiallyCorrelatedErrors2010} argue that unbounded variance inflation occurs in the posterior distribution of the fixed effects as the ratio of non-spatial to spatial variance approaches zero in a Bayesian linear model. To simplify the analysis of spatial smoothness in the multivariate BYM2 model of \eqref{eqn:spatial_analysis_model}, we let $R = rI_k$ such that $r \in (0, 1)$ represents the proportion of spatial variation, and define the smoothing ratio as $\frac{1 - r}{r}$. Spatial smoothness increases as the smoothing ratio approaches 0, or equivalently, $r \rightarrow 1^{-}$. We show that posterior variance inflation is limited in the multivariate BYM2 model of \eqref{eqn:spatial_analysis_model}. 

By spectral decomposition, there exist an orthogonal matrix $Q \in \mathbb{R}^{n \times n}$ and a diagonal matrix $K = \mbox{diag}(k_1, \ldots, k_n)$ with $k_i \geq 0$ for all $i = 1, \ldots, n$ such that $W_{\phi}^{1/2}(I - H) W_{\phi}^{1/2} = QKQ^{\T}$. Let $U = W^{1/2}_{\phi}Q$ and $K^{\star}$ be a $n \times n$ diagonal matrix such that $K^{\star}_{ii} = \mbox{I}(k_i > 0)$ for $i = 1, \ldots, n$, where $I(\cdot)$ is the indicator function. The derivations in Section~\ref{append:spatial_condpost_derivation} reveal 
\begin{equation}\label{eqn:spatial_beta_limit}
    \lim_{r \rightarrow 1^{-}} \Var(\vect(B_S) \given Y, M, R=rI_k, X_1) = (M^{\T}M) \otimes (X^{\T}X)^{-1}X^{\T}W_{\star}X(X^{\T}X)^{-1},
\end{equation}
where $W_{\star} = W_{\phi} - UK^{\star}U^{\T}$. Thus, the conditional posterior variance of the fixed effects $B_S$ cannot grow infinitely as the residual error becomes predominantly spatial. Similar calculations reveal $\lim_{r \rightarrow 1^{-}} \tilde{B}_S = \hat{B}_{NS}  - (X^{\T}X)^{-1}X^{\T}UK^{\star}U^{-1}Y$. 
The limiting between-outcome dependence is encoded by $M^{\T}M$, another potential source of variance inflation.

However, the limiting distribution of $M^{\T}M$ is also stable and depends on the structure of $W_{\phi}^{-1}$ through $U$ and $K^{\star}$. Section~\ref{append:limiting_distributions} derives the limiting distribution of $M^{\T}M$ as
\begin{equation}\label{eqn:spatial_M_limit}
    \lim_{r \rightarrow 1^{-}} \pi(M^{\T}M \given Y, R = rI_k, X_1) = \mbox{IW}(M^{\T}M \given v + n - p - 1, v\Sigma_{0} + Y^{\T}U^{-\T}K^{\star}U^{-1}Y),
\end{equation}
illuminating the effect of a minimal smoothing ratio on the scale of the conditional variance and between-outcome dependence in the multivariate BYM2 model. The asymptotic results of \eqref{eqn:spatial_beta_limit} and \eqref{eqn:spatial_M_limit} indicate that the posterior variance of the fixed effects in $B_S$ does not explode when the smoothing ratio is near zero, contrary to the case in \cite{hodgesAddingSpatiallyCorrelatedErrors2010}. 
Statistical inference remains intact under extreme spatial smoothing, unlike the pathological case of perfect multicollinearity in covariates, although \eqref{eqn:spatial_beta_limit} and \eqref{eqn:spatial_M_limit} assume $W_{\phi}^{-1}$ is invertible. This result aligns with 
\citet{wuAssessingSpatialDisparities2025a} who provide the limiting distribution of the total error variance as the smoothing ratio approaches zero in a univariate BYM2 model. 

\subsection{Unconditional and Conditional Association Estimands}\label{subsec:unconditional_versus_conditional}

We now consider the estimators $\tilde{B}_S$ and $\hat{B}_{NS}$ as point estimators derived from the Bayesian analysis models in Section~\ref{subsec:model_estimation} and consider their frequentist properties under the data generation model in \eqref{eqn:data_generating_model}. Although all parameters in \eqref{eqn:data_generating_model} are fixed, there is no universal consensus on the appropriate estimand for assessing the association between the covariates and response. For example, \cite{zimmermanDeconfoundingSpatialConfounding2022} and \cite{reichReviewSpatialCausal2021} consider the association between the covariates and response conditional on the confounder. In our setting, the parameter $B$ represents the association between $Y$ and $X$ conditional on $Z$. The conditional expectations of $\hat{B}_{NS}$ and $\tilde{B}_S$ under the data generation model in \eqref{eqn:data_generating_model} are $\Ep_{DG}[\hat{B}_{NS} \given Z, X_1] = B + (X^{\T}X)^{-1}X^{\T}Z$ and $\Ep_{DG}[\tilde{B}_{S} \given Z, X_1] = B + (X^{\T}X)^{-1}X^{\T}(Z - \Ep_{DG}[\hat{G} \given Z, X_1])$,
where $\Ep_{DG}[\vect(\hat{G}) \given Z, X_1] = (M^{\T} \otimes I_n)L^{-1}\vect((I_n - H)ZM^{-1})$. From this perspective, bias already exists in the non-spatial model from collinearity between $X$ and $Z$, and the spatial random effect adjusts the bias based on a product between $X$ and the projection of $Z$ onto the orthogonal complement of $X$ that is weighted according to both spatial structure and between-outcome covariance. The effect of this adjustment on the existing bias depends on the structure of $X$, $Z$, and the analysis parameters $M, R, W_{\phi}^{-1}$.

On the other hand, restricted spatial regression approaches such as those discussed in \cite{hodgesAddingSpatiallyCorrelatedErrors2010, hughesDimensionReductionAlleviation2013, hanksRestrictedSpatialRegression2015} target the unconditional association between the predictors and response, which is represented by $\Ep_{DG}[Y \given X_1] = X(B + D)$ in the multivariate BYM2 model, where $D = (\delta_{0}, D_1^{\T})^{\T}$ represents the linear dependency between the covariates $X_1$ and confounder $Z$ that is absorbed into the mean of $Y$. Taking the unconditional expectation of the spatial estimator in \eqref{eqn:spatial_beta_estimate} under the data generation model in \eqref{eqn:data_generating_model} reveals $\Ep_{DG}[\hat{B}_{NS}] = \Ep_{DG}[\tilde{B}_{S}] = B + D$, so both estimators are unbiased, regardless of the values of $M$, $R$, and $W_{\phi}^{-1}$ used to construct $\tilde{B}_{S}$.

Although we remain agnostic on this dichotomy, 
we focus on the estimation properties of $\hat{B}_{NS}$ and $\tilde{B}_S$ under the data generation model over repeated realizations of $(X_1, Z, Y)$. In the context of disease mapping, treating $X_1$ and $Z$ as stochastic is logical when observed and latent determinants of health, such as population characteristics or environmental factors, may dynamically change when measured alongside the health outcome or are estimated with finite precision. We next turn to whether the spatial analysis model improves estimation precision across multiple datasets of complete health outcome and disease risk maps when compared to the non-spatial analysis model.

\subsection{Estimation Over Multiple Datasets}\label{sec:multiple_datasets_estimation}


The spatial BYM2 model improves the mean-squared error of point estimates of $B$ over repeated realizations of $(X_1, Z, Y)$ if it correctly accounts for the variance in the predictors and confounding effects. 
Analysis in Section~\ref{append:spatial_variance} reveals that under the data generation model in \eqref{eqn:data_generating_model}, if $V_{\phi} = W_{\phi}$ and $\tilde{B}_{S} = \Ep[B_S \given Y, M = A, R = P, X_1]$, then 
\begin{equation}\label{eqn:spatial_efficiency}
    \Var_{DG}(\tilde{B}_{S,ij}) < \Var_{DG}(\hat{B}_{NS,ij}) \;\text{and}\; \Ep_{DG}\left[(\tilde{B}_{S,ij} - B_{ij})^2\right] < \Ep_{DG}\left[(\hat{B}_{NS,ij} - B_{ij})^2\right]
\end{equation}
for all $i = 1, \ldots, p + 1$ and $j = 1, \ldots, k$. In other words, if the variance structure of the spatial analysis model matches the data generation model and the conditional mean of $Z$ is a linear transformation of $X$, then $\tilde{B}_{S}$ is a more efficient estimator than $\hat{B}_{NS}$ in terms of mean squared error over repeated realizations of $(X_1, Y, Z)$. 
This result depends on correct specification of both the latent spatial variance structure and the correlation between outcomes. 


Uncertainty quantification in the spatial BYM2 model matches the variance of the point estimator under the data generation process. If $V_{\phi}^{-1} = W_{\phi}^{-1}$, the variance of $\tilde{B}_{S}$ in the data generation model conditional on $X_1$ is equal to the posterior variance in the analysis model,
\begin{equation}\label{eqn:variance_equality}
    \Var_{DG}(\vect(\tilde{B}_S) \given X_1) = \Var(\vect(B_S) \given Y, M = A, R = P, X_1),
\end{equation}
where the left-hand variance term is evaluated under the data generation model in \eqref{eqn:data_generating_model} and the right-hand term is evaluated under the spatial analysis model in \eqref{eqn:spatial_analysis_model} (see Section~\ref{append:spatial_variance} for a proof). 
Credible intervals for the unconditional estimand $F = B + D$ constructed using the conditional posterior will have coverage probability equal to the nominal coverage rate across realizations of $(Z, Y)$ with $X_1$ fixed. Combining \eqref{eqn:spatial_efficiency} and \eqref{eqn:variance_equality} implies that approaches such as restricted spatial regression methods that preserve the non-spatial estimates but reduce the posterior variance result in credible intervals that do not have a valid uncertainty interpretation under the data generation process. This aligns with previous results by \citet{zimmermanDeconfoundingSpatialConfounding2022, khanRestrictedSpatialRegression2022} that show restricted spatial regression methods tend to underestimate the true uncertainty in estimation of fixed effects. 

However, $\tilde{B}_{S}$ cannot be calculated if $A$ and $P$ are not known. For fully Bayesian inference, one solution is to use the marginal posterior mean of $B$ under \eqref{eqn:spatial_analysis_model}, which we denote as $\hat{B}_{S} = \Ep[B_{S} \given Y, X]$. 
Because $\hat{B}_{S}$ does not have a closed-form expression, we conduct simulation experiments to assess estimation performance and robustness of the analysis model when between-outcome dependence is unknown and spatial structure is misspecified.
 
\section{Simulation Analysis of Spatial Estimator}\label{sec:simulation}

We generate data sets $D^{(i)} = \left\{X_1^{(i)}, Y^{(i)}\right\}$ using \eqref{eqn:data_generating_model}, where $X_1^{(i)}$ is a $58 \times 1$ vector and $Y^{(i)}$ is a $58 \times 2$ matrix of $k = 2$ outcomes generated on a map of the $n = 58$ California counties for $i = 1, \ldots, 300$. 
We set $\beta_0 = (0, 2)^{\T}$, $B_1 = (1, 3)^{\T}$, $\delta_0 = (0.5, 0.5)^{\T}$, $D_1 = (0.3, 0.3)^{\T}$, $\mu = 0.5$, $A = \begin{bmatrix}
    1 & 0.5  \\
    0.5 & 1.0 \\ 
\end{bmatrix}$ such that $\Sigma = A^{\T}A = \begin{bmatrix}
    1.5 & 1.0 \\
    1.0 & 1.5 \\
\end{bmatrix}$, $P = \mbox{diag}(0.9, 0.7)$, and $C = 2$. We generate the data using a conditionally autoregressive (CAR) structure, $V^{-1}_\phi = c(D_W - \alpha W)$, where $W$ is the adjacency matrix, $D_W$ is a diagonal matrix with each diagonal element $D_{W, ii}$ being the number of neighbors of county $i$, $\alpha$ is the spatial smoothing parameter, and $c$ is a scaling factor such that the geometric mean of the diagonal elements of $W_\phi^{-1}$ is equal to one. We set $\alpha = 0.99$ and $c = 0.8352$ to construct $V_\phi^{-1}$ using the Californian county map.

 For the spatial analysis model in \eqref{eqn:spatial_analysis_model}, we employ the CAR prior through \eqref{eqn:full_spatial_prior} with $W_\phi = V_\phi$, and refer to this model as the unconditioned CAR model. Following \cite{jinOrderFreeCoRegionalizedAreal2007}, we set $v = 2$ for all simulation runs and $\Sigma_0$ as a diagonal matrix with diagonal entries equal to $||(I_n - H)Y_{j}||^2 / (n - p - 1)$. Although the prior $\pi(\Sigma_{S})$ is dependent on the data, \cite{jinOrderFreeCoRegionalizedAreal2007} argues that this enables robust posterior inference. We fix $\lambda_R = 0.01$ for $\pi(R)$, a vague prior on $R$ that attributes slightly higher prior density to smaller values of $r_i$. The spatial estimates $\hat{B}_{S}^{(i)}$ and 95\% highest posterior density credible intervals for each fixed effect are computed using 20,000 posterior samples after 20,000 burn-in samples obtained through a Metropolis within Gibbs sampling scheme detailed in Section~\ref{append:spatial_sampling_scheme} with proposal settings $s_1 = 0.10$, $s_2 = 0.15$, and $s_3 = 0.25$. For the conditioned CAR model, we calculate $\tilde{B}^{(i)}$ using \eqref{eqn:spatial_beta_estimate} by setting $M$, $R$, and $W_\phi$ equal to the true values of $A$, $P$, and $V_\phi$, respectively. 
 
 We compute 95\% conditional credible intervals for the elements of $F = B + D$ as $\tilde{B}_{ij} \pm z_{0.975}  \sqrt{\Var(B_{S, ij} \given Y, M = A, R = P, X_1)}$ for $i = 0, \ldots, p$ and $j = 1, \ldots, k$. For the non-spatial model in \eqref{eqn:nonspatial_analysis_model}, we set $\pi(B_{NS}, \Sigma_{NS}) \propto \mbox{IW}(\Sigma_{NS} \given v, v\Sigma_0)$ with the hyperparameters $v = 2$ and $v\Sigma_0$ identical to the scale matrix for $\pi(\Sigma_{S})$. We calculate 95\% highest posterior density credible intervals for each fixed effect using 10,000 exact posterior samples from $\pi(B_{NS}, \Sigma_{NS} \given Y, X_1)$. To examine the effects of misspecifying the spatial prior, we evaluate the aforementioned estimates and credible intervals under a SAR spatial prior on $G$ in \eqref{eqn:spatial_analysis_model} by setting $W_{\phi}^{-1} = c(I_n - \alpha \tilde{W})(I_n - \alpha\tilde{W})^{\T}$ where $\tilde{W}$ is a row-normalization of $W$ such that $\tilde{W}_{ij} = W_{ij} / \sum_{m = 1}^{n} W_{im}$, $\alpha = 0.99$ is the spatial smoothing parameter, and $c = 220.1809$ is a scaling factor such that the geometric mean of the prior marginal variances is equal to one.
 
\begin{table}[ht]
\centering
\begin{tabular}{cccccc}
  \hline
Frequentist Evaluation & Model & $F_{11}$ & $F_{12}$ & $F_{21}$ & $F_{22}$ \\ 
  \hline
\multirow{5}{*}{Mean-squared Error} & $B_{NS} \given Y, X_1$ & 0.642 & 0.592 & 0.112 & 0.109 \\ 
  & $B_{S} \given Y, X_1, A, P$; CAR Prior & 0.634 & 0.588 & 0.098 & 0.099 \\ 
  & $B_{S} \given Y, X_1, A, P$; SAR Prior & 0.635 & 0.598 & 0.103 & 0.104 \\ 
  & $B_{S} \given Y, X_1$; CAR Prior & 0.635 & 0.585 & 0.096 & 0.098 \\ 
  & $B_{S} \given Y, X_1$; SAR Prior & 0.674 & 0.612 & 0.101 & 0.103 \\ 
   \hline
\multirow{5}{*}{Coverage of $F = B + D$} & $B_{NS} \given Y, X_1$ & 0.357 & 0.370 & 0.657 & 0.713 \\ 
  & $B_{S} \given Y, X_1, A, P$; CAR Prior & 0.800 & 0.863 & 0.917 & 0.923 \\ 
  & $B_{S} \given Y, X_1, A, P$; SAR Prior & 0.930 & 0.967 & 0.837 & 0.847 \\ 
  & $B_{S} \given Y, X_1$; CAR Prior & 0.973 & 0.970 & 0.950 & 0.933 \\ 
  & $B_{S} \given Y, X_1$; SAR Prior & 0.990 & 0.997 & 0.613 & 0.690 \\ 
   \hline
\multirow{5}{*}{Average Posterior Variance} & $B_{NS} \given Y, X_1$ & 0.026 & 0.027 & 0.007 & 0.007 \\ 
  & $B_{S} \given Y, X_1, A, P$; CAR Prior & 0.185 & 0.221 & 0.008 & 0.009 \\ 
  & $B_{S} \given Y, X_1, A, P$; SAR Prior & 0.417 & 0.476 & 0.007 & 0.008 \\ 
  & $B_{S} \given Y, X_1$; CAR Prior & 0.480 & 0.420 & 0.008 & 0.009 \\ 
  & $B_{S} \given Y, X_1$; SAR Prior & 0.927 & 0.808 & 0.002 & 0.004 \\ 
   \hline
\end{tabular}

\caption{Frequentist evaluation of point estimates and 95\% credible intervals for $F = B + D$ from non-spatial/spatial CAR and SAR models across 300 datasets generated under a CAR spatial data generation model on a map of Californian counties.} 
\label{tab:sim_estimation_results}
\end{table}
The frequentist evaluation of the posterior means and 95\% credible intervals for each analysis model in Table~\ref{tab:sim_estimation_results} provides simulation evidence for the claim of \eqref{eqn:spatial_efficiency}, as the conditional spatial estimator $\tilde{B}_{S}$ is superior to $\hat{B}_{NS}$ in terms of mean squared error for every coefficient despite the presence of the spatial confounder $Z$ in the data generation model. The spatial models show inflated posterior variances for the intercept terms compared to the non-spatial model. The conditioned and unconditioned CAR model have similar mean-squared errors, while the credible intervals of the conditioned CAR model exhibit modest undercoverage. The unconditioned spatial CAR prior provides frequentist coverage closest to the nominal credible level, while the non-spatial model does not yield a coverage rate above 75\% for any coefficient. 
These results illustrate that the multivariate spatial analysis model in \eqref{eqn:spatial_analysis_model} can improve estimation compared to the non-spatial analysis model in \eqref{eqn:nonspatial_analysis_model} despite the effects of spatial confounding and uncertainty about the variance parameters $A$ and $P$. Increased posterior variance in the spatial model compared to the non-spatial model is not a inherent flaw, but rather a reason to distrust the non-spatial model which suffers from undercoverage. 

 For the SAR analysis, the efficiency result in \eqref{eqn:spatial_efficiency} is not guaranteed as the analysis spatial structure $W_{\phi}^{-1}$ does not match $V_{\phi}^{-1}$ in the true data generating model. Nonetheless, the unconditioned spatial model obtains lower mean squared errors and credible interval coverage rates closer to the nominal level compared to the non-spatial model (Table~\ref{tab:sim_estimation_results}). Therefore, a confounded and misspecified spatial effect may still be preferable to none for estimation, albeit at the cost of large posterior variance inflation for the intercept terms and potentially inaccurate uncertainty quantification. In practice, we recommend fitting multiple spatial models and using an information criteria such as DIC \citep{spiegelhalterBayesianMeasuresModel2002} or WAIC \citep{watanabeAsymptoticEquivalenceBayes2010} to perform model selection. 

\begin{figure}[!t]
    \centerline{\includegraphics[width=0.8\linewidth]{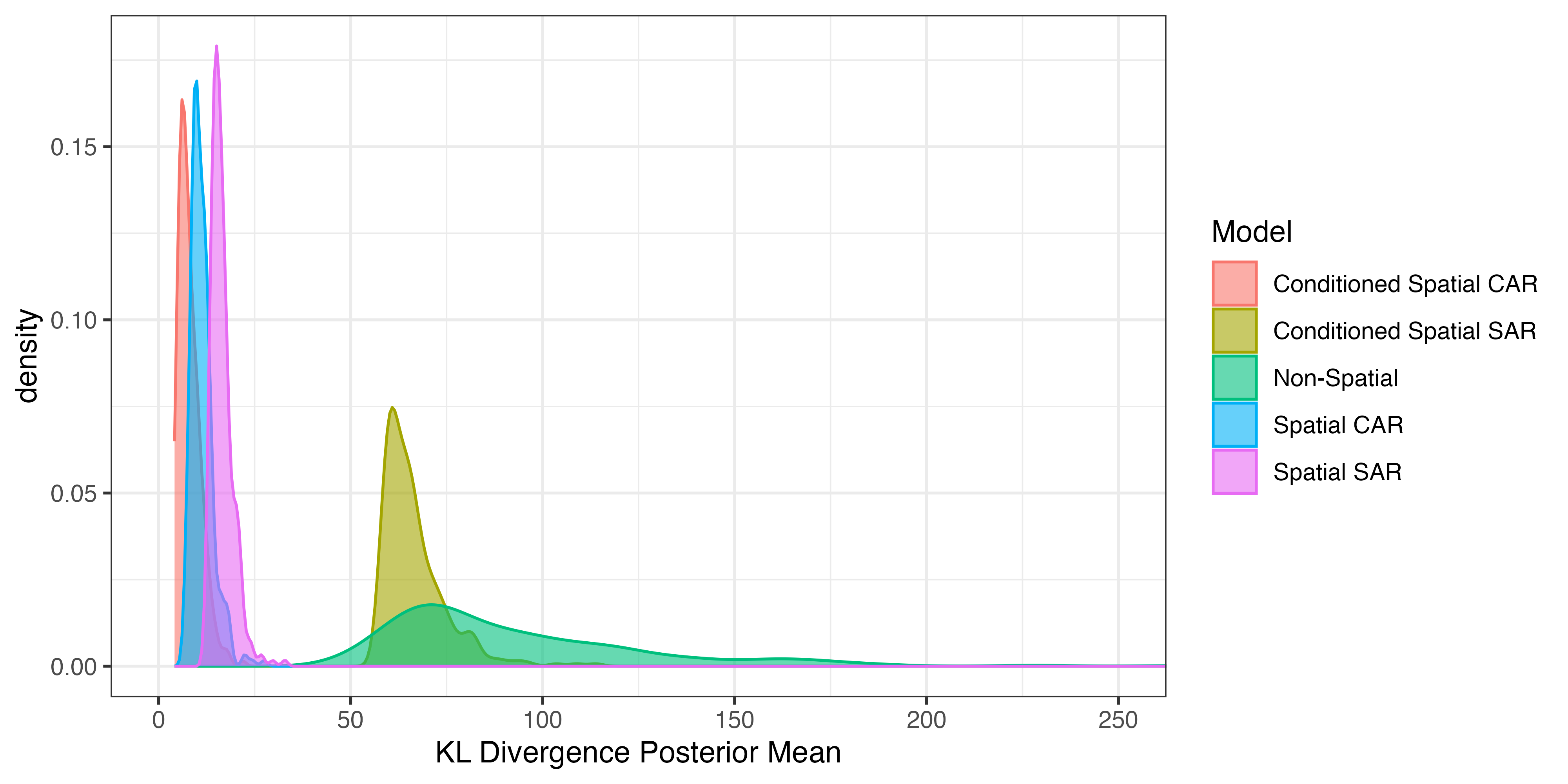}}
    \caption{Density plot of posterior expectations of $D_{KL}(p(Y_0) || q(Y_0))$ for the non-spatial model, conditioned CAR and SAR spatial models, and unconditioned CAR and SAR spatial models over 300 datasets simulated from a CAR data generation model.}
    \label{fig:sim_KL_plot}
\end{figure}

We assess model fit by computing the Kullback-Leiber divergence $D_{KL}(p(Y_0) || q(Y_0))$ for each dataset and analysis model. Here, $Y_0$ is an observation from a new realization of the underlying stochastic process conditional on $X_1$, $p(Y_0)$ is the data generation density, and $q(Y_0)$ is the predictive density under a analysis model. Marginalizing out $Z$ in the data generation model from \eqref{eqn:data_generating_model} yields $p(Y_0) = \mbox{N}_{nk}(\vect(Y_0) \given X(B + D), A^{\T}(I_k - P)A \otimes I_n + A^{\T}PA \otimes V_{\phi})$. In the non-spatial model, $q(Y_0) = \mbox{N}_{nk}(\vect(Y_0) \given \vect(XB_{NS}), \Sigma_{NS} \otimes I_n)$. For the conditioned spatial model, $q(Y_0) = \mbox{N}_{nk}(\vect(Y_0) \given \vect(XB_{S}), A^{\T}(I_k - P)A \otimes I_n + A^{\T}PA \otimes W_{\phi})$, and for the unconditioned spatial model, $q(Y_0 \given X_{1}) = \mbox{N}_{nk}(\vect(Y_0) \given \vect(XB_{S}), M^{\T}(I_k - R)M \otimes I_n + M^{\T}RM \otimes W_{\phi})$.  We use the aforementioned posterior samples for the non-spatial and unconditioned spatial model along with 20,000 exact samples of $B_{S}$ in the conditioned spatial model to obtain samples from the posterior distribution of $D_{KL}(p(Y_0) || q(Y_0))$ for each dataset and analysis model. The density plot of the posterior means in Figure~\ref{fig:sim_KL_plot} demonstrates that the unconditioned spatial models generally provide a better fit than the non-spatial model. Therefore, the increased posterior variance in the spatial models is a more accurate representation of the true variability in the data generation model. The unconditioned SAR model outperforms the conditioned SAR model, adapting to the misspecified spatial structure.

\section{Obesity, Diabetes, and Diabetes-Related Cancer Mortality in US Counties}\label{sec:application}


We conduct a spatial and non-spatial analysis of US county-level associations between structural characteristics and three health outcomes: obesity prevalence, diabetes prevalence, and diabetes-related cancer mortality. Obesity and diabetes are associated with increased risk of cancer, treatment complications, and higher mortality rate \citep{harborgNewHorizonsEpidemiology2024}. However, many epidemiological investigations often study relations between socioeconomic or environmental factors and a single health outcome such as obesity prevalence \citep[see, e.g.,][ and references therein]{slackGeographicConcentrationUs2014, bennettObesityWorkingAge2011, salernoCountylevelSocioenvironmentalFactors2024}, diabetes prevalence \citep[e.g.,][]{hill-briggsSocialDeterminantsHealth2020, cunninghamCountylevelContextualFactors2018}, or cancer mortality rates \citep[e.g.,][]{oconnorFactorsAssociatedCancer2018, 
dongVariationFactorsAssociated2022}. Furthermore, previous epidemiological studies have demonstrated the need to account for spatial autocorrelation on the county geographic level \citep[e.g.,][]{ohFoodDesertsExposure2024, slackGeographicConcentrationUs2014, hippSpatialAnalysisCorrelates2015}. We seek to consolidate the aforementioned analysis goals while accounting for spatial autocorrelation and more efficiently estimating associations according to the results in Section \ref{sec:multiple_datasets_estimation}. In Section~\ref{append:RDA_setup}, we contrast our results to an analysis using the non-spatial model in \eqref{eqn:nonspatial_analysis_model}.
\begin{figure}[!t]
    \centerline{\includegraphics[width=0.8\linewidth]{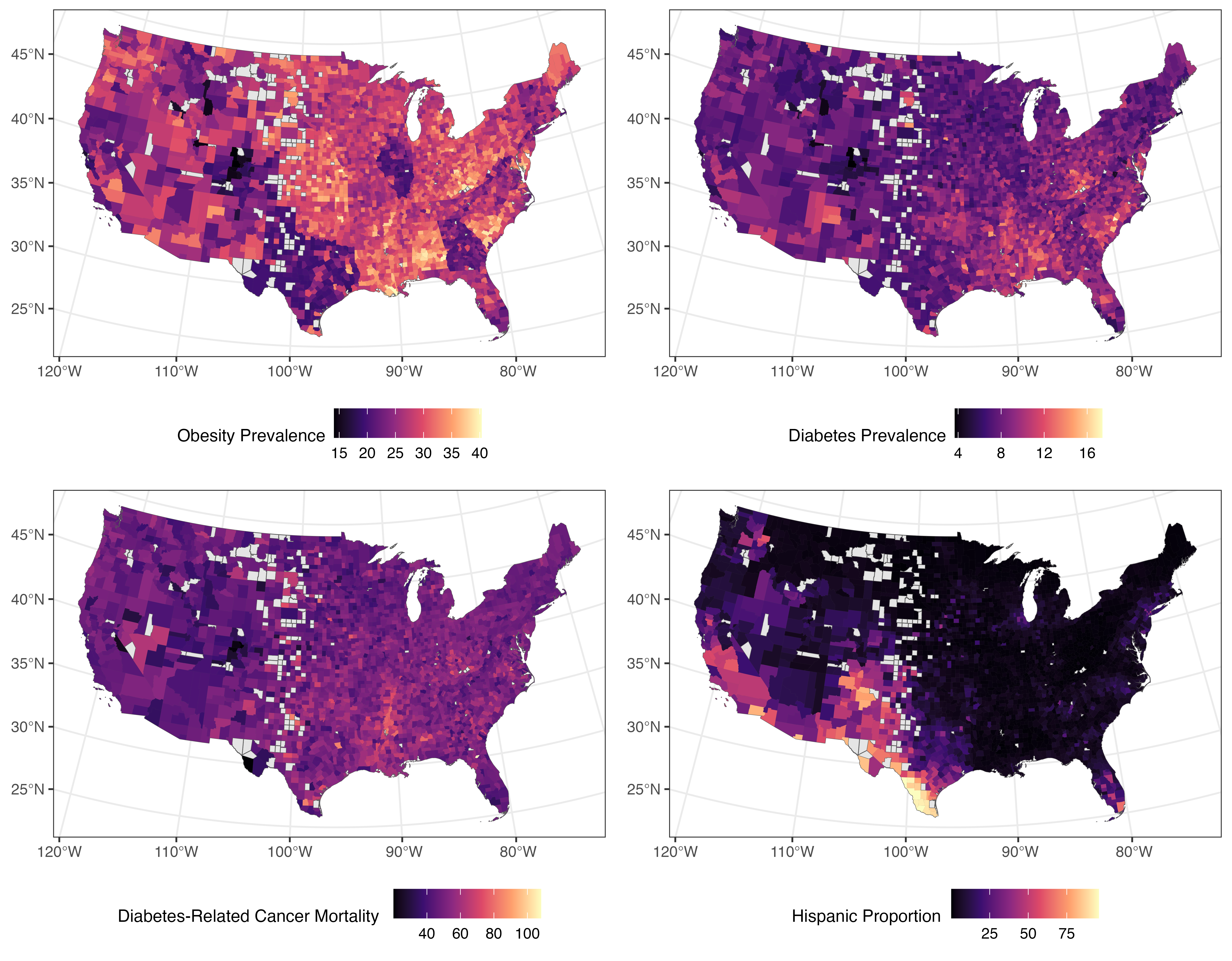}}
    \caption{Estimates of US county-level obesity prevalence in 2015, diabetes prevalence in 2015, diabetes-related cancer mortality rates from 2010 to 2020, and percentage of residents that were Hispanic in 2010. Counties with a missing covariate or response are shown in gray and excluded from the analysis.}
    \label{fig:data_maps}
\end{figure}

 Our analysis accounts for county-level characteristics across five established critical determinants of population health, which include the economic context, healthcare context, natural and built environment, educational context, and population structure \citep{
slackGeographicConcentrationUs2014, myersRegionalDisparitiesObesity2015, myersChangeObesityPrevalence2016}. For brevity, we refer to Section~\ref{append:RDA_setup} for the original data sources and descriptions of the response and covariate data. After aggregating all variables into one dataset, we drop counties with missing covariate data and subset to a contiguous region consisting of $n = 2,960$ US counties for simplicity of analysis. For subsequent analysis, we center and scale the data such that each outcome and covariate has sample mean and standard deviation equal to zero and one, respectively. The three health outcomes as well as county proportions of Hispanic residents are mapped on the final study region in Figure~\ref{fig:data_maps}, indicating the presence of considerable spatial clustering, such as the high obesity prevalence counties in the East North Central and East South Central states.  

We first assess spatial autocorrelation by computing Moran's $I$ and Geary's $C$ using the residuals of an ordinary least squares linear model for each outcome \citep{banerjeeHierarchicalModelingAnalysis2015}. Each statistic is significant under 10,000 random permutations (p-value = 1/10,001), indicating significant spatial dependence in 
all three health outcomes. To implement \eqref{eqn:spatial_analysis_model}, we set the joint prior as in \eqref{eqn:full_spatial_prior} with $v = 3$, $\lambda_R = 0.01$, and $\Sigma_0$ as a diagonal matrix where the $j$th diagonal element is $||(I_n - H)Y_j||^2 / (n - p - 1)$ following the prior recommendations of \cite{jinOrderFreeCoRegionalizedAreal2007}. We place a spatial CAR prior on the random effects by setting $W_\phi^{-1} = c(D_W - \alpha W)$ with the neighbor and adjacency matrices of the study region, $\alpha = 0.99$, and $c = 0.3963$. This CAR prior results in a better fit quantified by a lower DIC compared to a SAR spatial prior. 
After standardizing all responses and predictors to have mean zero and unit variance, we employ the Metropolis within Gibbs sampling scheme detailed in Section~\ref{append:spatial_sampling_scheme} with proposal hyperparameters $s_1 = 0.05$, $s_2 = 0.05$, and $s_3 = 0.45$ to obtain $10,000 \times 4\;\text{chains} = 40,000$ posterior samples of $\{B_S, \gamma, M, R\}$. One sample is saved every five update loops to 
conserve computer memory. MCMC convergence is assessed via the $\widehat{R}$ and effective sample size statistics computed using the \texttt{rstan} package in R. The $\widehat{R}$ statistic is below the recommended threshold of 1.01 and the effective sample size exceeds 1000 for all sampled parameters  \citep{vehtariRankNormalizationFoldingLocalization2021}. Coefficients are deemed statistically significant if the corresponding 95\% highest posterior density credible interval does not contain zero.


\begin{table}[ht]
\centering
\resizebox{\columnwidth}{!}{\begin{tabular}{llll}
  \hline
Variable & Diabetes & Obesity & Cancer Mortality \\ 
  \hline
Intercept & 1.8e-03 (-0.12, 0.12) & 2.2e-05 (-0.11, 0.1) & -8e-03 (-0.17, 0.14) \\ 
  Poverty Rate & 0.031 (-0.039, 0.098) & -0.058 (-0.12, 2.5e-03) & -0.043 (-0.13, 0.042) \\ 
  Median Income & -0.093 (-0.16, -0.032) & -0.15 (-0.21, -0.097) & -0.18 (-0.26, -0.099) \\ 
  Unemployment & 0.061 (0.017, 0.11) & 9.4e-03 (-3e-02, 0.047) & -0.053 (-0.11, 3.5e-03) \\ 
  Uninsured Rate & -0.011 (-0.067, 0.045) & -0.18 (-0.23, -0.13) & -0.1 (-0.17, -3e-02) \\ 
  PCP Density & -0.022 (-0.054, 0.011) & -0.045 (-0.074, -0.017) & -3e-02 (-0.071, 0.011) \\ 
  Outpatient Visits & 3e-02 (1.4e-03, 0.057) & 0.024 (-1.4e-04, 0.048) & 5.4e-03 (-0.029, 0.041) \\ 
  Low Access & 0.024 (-1.2e-03, 0.052) & 0.018 (-4.6e-03, 0.042) & 0.024 (-9.1e-03, 0.057) \\ 
  SNAP Assistance & 0.089 (0.028, 0.15) & 0.029 (-0.027, 0.083) & 0.32 (0.25, 0.4) \\ 
  Physical Inactivity & 0.47 (0.44, 0.5) & 0.47 (0.44, 0.5) & 0.046 (8.3e-03, 0.085) \\ 
  Recreation Facilities & -4e-02 (-0.068, -0.012) & -0.011 (-0.035, 0.014) & -0.015 (-5e-02, 2e-02) \\ 
  Percent NH-Black & 0.22 (0.17, 0.27) & 0.11 (7e-02, 0.16) & 0.15 (0.084, 0.21) \\ 
  Percent Hispanic & -0.052 (-0.11, 5.8e-03) & -0.081 (-0.13, -3e-02) & -0.15 (-0.22, -0.074) \\ 
  Percent $\geq$ 65 years & -0.091 (-0.13, -0.049) & -0.064 (-0.1, -0.027) & 0.073 (0.021, 0.13) \\ 
  Percent $\leq$ 18 years & 0.084 (0.044, 0.13) & 0.094 (0.058, 0.13) & 0.088 (0.035, 0.14) \\ 
  Urban Percent & 0.085 (0.046, 0.12) & 0.069 (0.036, 0.1) & 0.061 (0.012, 0.11) \\ 
  HS Diploma Rate & 0.047 (-9.8e-03, 0.1) & -0.017 (-0.066, 0.033) & -0.12 (-0.19, -0.052) \\ 
   \hline
\end{tabular}}
\caption{Spatial model posterior means and 95\% highest posterior density credible intervals for regression of US counties' diabetes prevalence, obesity prevalence, and diabetes-related cancer mortality rates in 2015.} 
\label{tab:spatial_model_estimates}
\end{table}


Table~\ref{tab:spatial_model_estimates} lists the standardized coefficient estimates for the spatial analysis model, indicating that diabetes prevalence in US counties has a moderate positive association with higher percentages of adults that are physically inactive and a weak association with percent of non-Hispanic black residents. Other significant predictors of diabetes prevalence include median income, unemployment rate, outpatient visit rate, percentage SNAP recipients, density of recreation facilities, percentage of residents at least 65 years of age, percentage of residents at most 18 years of age, and percentage of residents in urban areas. 
County obesity prevalence rates correlates positively with higher rates of physical inactivity, higher proportion of non-Hispanic black residents, lower median income, and uninsured rate. Other significant independent variables weakly associated with higher obesity prevalence include density of primary care physicians, percentage of Hispanic residents, percentage of residents at least 65 years of age, percentage of residents at most 18 years of age, and percentage of residents in urban areas. 
Higher diabetes-related cancer mortality rates are weakly associated with percentage of SNAP recipients, median income, rates of high school diploma attainment, percentage of non-Hispanic black residents, percentage of Hispanic residents, percentage of residents without health insurance, percentage of adults that are physically inactive, percentage of residents at least 65 years of age, percentage of residents at most 18 years of age, and percentage of residents in urban areas.

\section{Discussion}\label{sec:discussion}

This manuscript formally investigates spatial confounding in the presence of multivariate dependencies among outcomes. Our core takeaway is that a spatial model afflicted by spatial confounding may still produce more efficient estimates with accurate uncertainty quantification compared to a non-spatial model. Adding a spatial effect may alter significant associations, but is not sufficient reason to discard the spatial model. Our simulation results also reveal that the spatial analysis model is robust to misspecified spatial structures. 

To our knowledge, we are the first to use a multivariate spatial model to estimate county-level associations for health outcomes of diabetes, obesity, and cancer in the US simultaneously. Although the set of significant predictors differs between spatial and non-spatial models, the resulting credible intervals do not indicate systematic variance inflation from spatial confounding. Previous non-spatial analysis \citep{knightMultimethodMultidatasetAnalysis2020, hippSpatialAnalysisCorrelates2015} indicates that Hispanics are at higher risk of diabetes, but counties with a higher proportion of Hispanic residents are associated with lower rates of diabetes prevalence. Our analysis indicates that the county-level association may be an artifact of spatial clustering of counties with large Hispanic populations. However, large percentages of Hispanic populations are associated with a decrease in obesity prevalence and diabetes-related cancer mortality rates after accounting for spatial dependence. More research is needed to understand the disparity in the levels of Hispanic association at the individual and county levels. 

Although exposure to food deserts was not a significant predictor of any of the three health outcomes after accounting for other county-level characteristics in the spatial model, the county-level percentage of SNAP recipients was a significant predictor of diabetes-related cancer mortality with the largest magnitude standardized coefficient estimate in the spatial model. This result is consistent with those of \cite{chenCountylevelFoodInsecurity2023} and \cite{hongAssociationFoodInsecurity2024}, which warrant further investigation into the potential causal role that food insecurity plays in related cancer treatment outcomes at the individual and county level. 
\cite{ohFoodDesertsExposure2024} reports lack of a global association between food environment measures and county-level diabetes or obesity rates, but demonstrates that methods such as geographically weighted regression can identify specific counties that may benefit from regional initiatives such as reducing unhealthy food outlets, introducing more supermarkets or improving park access.

\section{Supplemental Materials}

Supplementary materials include theoretical derivations, sampling algorithm details, details on data sources and variables in Section~\ref{sec:application}, and a non-spatial analysis of the obesity/diabetes prevalence and diabetes-related cancer mortality data.  All computer programs required to reproduce the simulation and data analysis in this manuscript are available at \url{https://github.com/Ky-Wu/spatial-confounding}.

\bibliography{spatial_confounding}
\bibliographystyle{abbrvnat}

\end{document}


\maketitle

\section{Conditional Posterior of Fixed Effects in Multivariate BYM2}\label{append:spatial_condpost_derivation}

Let $M_0 \in \mathbb{R}^{(p + 1)k \times (p + 1)k}$ be positive definite, $m_0 \in \mathbb{R}^{n(p + 1) \times 1}$, and $W_\phi \in \mathbb{R}^{n \times n}$ be positive definite. If $\pi(B_{S}, G, M, R) = N_{pk}\left(\vect(B_S) \given M_0m_0, M_0\right) \times \mbox{N}_{nk}\left(\vect(G) \given 0_{nk}, M^{\T}RM \otimes W_\phi\right) \times \pi(M, R)$ is the prior for \eqref{eqn:spatial_analysis_model}, then the joint posterior distribution is equivalent to that of the augmented model \citep{banerjeeModelingMassiveSpatial2020b},
\begin{equation*}\label{eqn:augmented_system}
    \underbrace{\begin{bmatrix}
        \vect(Y) \\
        M_0m_0 \\
        0_{nk} 
    \end{bmatrix}}_{y_\star} =
    \underbrace{\begin{bmatrix}
        I_k \otimes X & I_{nk} \\
        I_{(p + 1)k} & 0_{(p + 1)k \times nk} \\
        0_{nk \times (p + 1)k} & I_{nk}
    \end{bmatrix}}_{X_\star}
    \underbrace{\begin{bmatrix}
        \vect(B_{S}) \\
        \vect(G)
    \end{bmatrix}}_{B_\star}
    + \underbrace{\begin{bmatrix}
        \eta \\
        \eta_{B_{S}} \\
        \eta_{\phi}
    \end{bmatrix}}_{\eta_{\star}}, \;\;
    \eta_{\star} \sim N_{k(2n + p + 1)}\left(0_{k(2n + p + 1)}, V_{y_\star}\right)\;,
\end{equation*}
where $V_{y_\star} = \begin{bmatrix}
        M^{\T}(I_k - R)M \otimes I_n & 0_{nk \times (p + 1)k} & 0_{nk \times nk} \\
        0_{(p + 1)k \times nk} & M_0 & 0_{(p + 1)k \times nk} \\
        0_{nk \times nk} & 0_{nk \times (p + 1)k} & M^{\T}RM \otimes W_\phi
    \end{bmatrix}$ and the prior $\pi(B_{\star}, M, R) = \pi(M, R)$. Then, $\pi(B_{\star} \given y_{\star}, M, R, X_1) = N_{(n + p + 1)k}\left(B_{\star} \given F_{B_{\star}} f_{B_{\star}}, F_{B_{\star}}\right)$ follows from standard Bayesian linear regression computations where $f_{B_{\star}} = X_{\star}^{\T}V_{y_\star}^{-1}y_{\star}$ and $F_{B_{\star}} = (X_{\star}^{\T}V_{y_\star}^{-1}X_{\star})^{-1}$. In the case when $\pi(B_{S}, G, M, R) \propto \mbox{N}_{nk}\left(\vect(G) \given 0_{nk}, M^{\T}RM \otimes W_\phi\right) \times \pi(M, R)$, 
\begin{equation}\label{eqn:beta_star_postparam}
\begin{split}
      f_{B_{\star}} = \begin{bmatrix}
        \vect(X^{\T}YM^{-1}(I_k - R)^{-1}M^{-\T}) \\
        \vect(YM^{-1}(I_k - R)^{-1}M^{-\T})
    \end{bmatrix},\\ F_{B_{\star}} = \begin{bmatrix}
        M^{\T}(I_k - R)M \otimes (X^{\T}X)^{-1} + F_{1}F_{2}F^{\T}_{1} & - F_{1}F_{2} \\
        -F_{2}F_{1}^{\T} & F_2
    \end{bmatrix},
\end{split}
\end{equation}
where $F_1 = I_k \otimes (X^{\T}X)^{-1}X^{\T}$ and $F_2^{-1} = M^{-1}(I_k - R)^{-1}M^{-\T} \otimes (I - H) + M^{-1}R^{-1}M^{-\T} \otimes W_{\phi}^{-1}$. Evaluating $F_{B_{\star}}f_{B_{\star}} = (X_{\star}V_{y_\star}^{-1}X_{\star})^{-1}X_{\star}V_{y_\star}^{-1}y_{\star}$ results in the expressions for $\Ep[B_{S} \given Y, M, R, X_1]$ and $\Ep[\vect(G) \given Y, M, R, X_1]$ in \eqref{eqn:spatial_beta_estimate}.

The upper-left block element of $F_{B_{\star}}$ is the conditional variance $\Var(B_S \given Y, M, R, X_1)$. Since $W_{\phi}^{-1}$ is assumed to be positive definite and $I_n - H$ is positive semi-definite, by spectral decomposition, there exists an orthogonal matrix $Q$ and diagonal matrix $K = \mbox{diag}(k_1, \ldots, k_n)$ with $k_i \geq 0$ for all $i$ such that $W_{\phi}^{1/2}(I - H)W_{\phi}^{1/2} = QKQ^{\T}$. Set $U = W_{\phi}^{1/2}Q$, then $U$ is invertible, $W_{\phi}^{-1} = U^{-\T}U^{-1}$, $I_n - H = U^{-\T}KU^{-1}$, and the conditional posterior variance of $B_S$ can be expressed as
\begin{equation}\label{eqn:postvar_cond_spatial_estimator}
\begin{split}
    \Var(\vect(B_S) \given Y, M, R, X_1) =& M^{\T}(I_k - R)M \otimes (X^{\T}X)^{-1} \\ &+ S_1\left((I_k - R)^{-1} \otimes K + R^{-1} \otimes I_n \right)^{-1}S_1^{\T},
\end{split}
\end{equation}
where $S_1 = M^{\T} \otimes (X^{\T}X)^{-1}X^{\T}U$. Letting $R = rI_k$, for any $d_i$, $\lim_{r \rightarrow 1^{-}} \frac{1}{1/ r + k_i / (1 - r)} = \mbox{I}(k_i = 0)$. Let $K^{\star}$ be a $n \times n$ diagonal matrix such that $K^{\star}_{ii} = \mbox{I}(k_i > 0)$ for $i = 1, \ldots, n$. Then, $\lim_{r \rightarrow 1^{-}} \left((I_k - R)^{-1} \otimes K + R^{-1} \otimes I_n \right)^{-1} = I_k \otimes (I_n - K^{\star})$ and the posterior conditional variance approaches the limit
\begin{equation*}
    \lim_{r \rightarrow 1^{-}} \Var(\vect(B_S) \given Y, M, R = rI_k, X_1) = (M^{\T}M) \otimes (X^{\T}X)^{-1}X^{\T}W_{\star}X(X^{\T}X)^{-1},
\end{equation*}
where $W_{\star} = W_{\phi} - UK^{\star}U^{\T}$.

\section{Limiting Posterior Distribution of $M^\T M$ in Multivariate BYM2 Model}\label{append:limiting_distributions}

Under the spatial analysis model of \eqref{eqn:spatial_analysis_model}, let the prior be given by \eqref{eqn:full_spatial_prior}. We show that the limit of $\pi(M^{\T}M \given Y, R = rI_k, X_1)$ as $r \rightarrow 1^{-}$ is equal to the result in \eqref{eqn:spatial_M_limit}. Set $B_{\star} = (\vect(B_{S})^{\T}, \vect(G)^{\T})^{\T}$, and let $f_{B_{\star}}$ and $F_{B_{\star}}$ be defined as in \eqref{eqn:beta_star_postparam}. By the use of an augmented model as in Appendix~\ref{append:spatial_condpost_derivation}, 
\begin{align*}
    \pi(&B_{\star}, M \given Y, R, X_1) \propto \pi(Y \given B_{\star}, M, R, X_1) \pi(B_{\star}, M\given R, X_1) \\
    &\propto \exp\left(-\frac{1}{2}\left(B_{\star}^{\T}F_{B_{\star}}^{-1}B_{\star} - 2B_{\star}^{\T}f_{B_{\star}} + y_{\star}^{\T}V_{y_{\star}}^{-1}y_{\star}\right)\right)|M^{\T}M|^{-n}\times \pi(M) \\
    &\propto N_{(n + p + 1)k}(B_{\star} \given F_{B_{\star}}f_{B_{\star}}, F_{B_{\star}}) \times \exp\left(-\frac{1}{2}\left(y_{\star}^{\T}V_{y_{\star}}^{-1}y_{\star} - f_{B_{\star}}^{\T}F_{B_{\star}}f_{B_{\star}}\right)\right) |M^{\T}M|^{-\frac{n - p - 1}{2}}\times \pi(M). 
\end{align*}
Hence, $\pi(M^{\T}M \given Y, R, X_1) \propto \exp\left(-\frac{1}{2}\left(y_{\star}^{\T}V_{y_{\star}}^{-1}y_{\star} - f_{B_{\star}}^{\T}F_{B_{\star}}f_{B_{\star}}\right)\right) |M^{\T}M|^{-\frac{n - p - 1}{2}}\times \pi(M^{\T}M)$. Matrix calculations reveals that $y_{\star}^{\T}V_{y_{\star}}^{-1}y_{\star} - f_{B_{\star}}^{\T}F_{B_{\star}}f_{B_{\star}} = \vect(YM^{-1})^{\T}Q_1 \vect(YM^{-1})$, where
\begin{equation*}
    Q_1 = (I_k - R)^{-1} \otimes (I_n - H) - \left((I_k - R)^{-1} \otimes (I_n - H)\right)B^{-1}_{\dagger}(I_k \otimes (I_n - H))
\end{equation*}
and $B_{\dagger} = I_k \otimes (I_n - H) + R^{-1}(I_k - R)\otimes W_{\phi}^{-1}$. By spectral decomposition, there exists an orthogonal matrix $Q$ and diagonal matrix $K = \mbox{diag}(k_1, \ldots, k_n)$ with $k_i \geq 0$ for all $i$ such that $W_{\phi}^{1/2}(I - H)V_{\phi}^{1/2} = QKQ^{\T}$. Set $U = W_{\phi}^{1/2}Q$, then $I_n - H = U^{-\T}KU^{-1}$, $W_{\phi}^{-1} = U^{-\T}U^{-1}$, and $Q_1 = (I_k \otimes U^{-\T})Q_2(I_k \otimes U^{-1})$, where
\begin{equation*}
    Q_2 = (I_k - R)^{-1} \otimes K - ((I_k - R)^{-1} \otimes K^2)\left(I_k \otimes K + R^{-1}(I_k - R) \otimes I_n\right)^{-1}
\end{equation*}
Let $R_{ii} = r \in (0, 1)$ for all $i = 1, \ldots, k$. For any given $k_j$, $\frac{k_j}{1 - r} - \frac{k_j^2}{(1 - r)(k_j + (1 - r)/r)} = \frac{k_j}{rk_j + 1 - r}$, which converges to $0$ as $r \rightarrow 1^{-}$ if $k_j = 0$ and converges to $1$ if $k_j > 0$. Therefore, $\lim_{r \rightarrow 1^{-}} Q_1 = I_k \otimes U^{-\T}K^{\star}U^{-1}$, where $K^{\star}$ is a $n \times n$ diagonal matrix with $K^{\star}_{ii} = \mbox{I}(k_i > 0)$ for all $i = 1, \ldots, n$. Further simplification shows that $\lim_{r \rightarrow 1^{-}} (y_{\star}^{\T}V_{y_{\star}}^{-1}y_{\star} - f_{B_{\star}}^{\T}F_{B_{\star}}f_{B_{\star}}) = \tr\left(Y^{\T}U^{-\T}K^{\star}U^{-1}YM^{-1}M^{-\T}\right)$. The final result is 
\begin{equation*}
    \lim_{r \rightarrow 1^{-}} \pi(M^{\T}M \given Y, r, X_1) \propto |M^{\T}M|^{-\frac{v +n - p - 1}{2}}\exp\left(-\frac{1}{2} \tr\left((v\Sigma_{0} + Y^{\T}U^{-\T}K^{\star}U^{-1}Y)M^{-1}M^{-\T}\right)\right) 
\end{equation*}
which identifies as the density of an inverse-Wishart distribution with parameters given by \eqref{eqn:spatial_M_limit}.

\section{Data Generation Variance of Bayesian Estimators}\label{append:spatial_variance}

Suppose that under the data-generating model, $\Var(\vect(Y) \given X_1) = A^{\T}PA \otimes V_{\phi} + A^{\T}(I_k - P)A \otimes I_n$, where $V_{\phi}$ is a fixed $n \times n$ positive definite matrix capturing the true spatial covariance structure of the latent factors, $A$ is an invertible $k \times k$ matrix, and $P = \mbox{diag}(\rho_{1}, \ldots, \rho_{k})$ with $\rho_{i} \in (0, 1)$ for $i = 1, \ldots, k$ (such as in the data-generation model discussed in Section \ref{sec:dg_analysis_models}). Consider the analysis model in \eqref{eqn:spatial_analysis_model} with the working spatial covariance matrix $W_{\phi}$ equal to the data generation spatial covariance matrix $V_\phi$, $\tilde{B}_{S}= \Ep[B_s \given Y, M = A, R = P, X_1]$, and $\hat{G} = \Ep[G \given Y, M = A, R = P, X_1]$. We show that $\Var(\tilde{B}_{S,ij} \given X_1) \leq \Var(\hat{B}_{NS,ij} \given X_1)$ for $i = 0, \ldots, p$ and $j = 1, \ldots, k$. Under the data-generation model, 
\begin{equation*}
\begin{split}
    \Cov_{DG}(\vect(\hat{B}_{NS}), \vect((X^{\T}X)^{-1}X^{\T}\hat{G}) \given X_1) = (A^{\T}P &\otimes (X^{\T}X)^{-1}X^{\T}V_{\phi}(I_n - H)) \\
    &\times B^{-1}_{\dagger}(A \otimes X(X^{\T}X)^{-1}) 
\end{split}
\end{equation*}
since $X^{\T}(I - H) = 0$. Since $V_{\phi}^{-1}$ is assumed to be positive definite and $I_n - H$ is positive semi-definite, then by spectral decomposition, there exists an orthogonal matrix $Q$ and diagonal matrix $K = \mbox{diag}(k_1, \ldots, k_n)$ with $k_i \geq 0$ for all $i$ such that $V_{\phi}^{1/2}(I - H)V_{\phi}^{1/2} = QKQ^{\T}$. Set $U = V_{\phi}^{1/2}Q$, then $U$ is invertible, $U^{\T}V_{\phi}^{-1}U = I_n$, $U^{\T} (I_n - H)U = K$, $V_{\phi}(I_n - H) = UKU^{-1}$, $(I_n - H)V_{\phi}(I_n - H) = U^{-\T}K^2U^{-1}$, and $B^{-1}_{\dagger} = (I_k \otimes U)\left(I_k \otimes K + P^{-1}(I_k - P) \otimes I_n\right)^{-1}(I_k \otimes U^{\T})$. This means that $(P\otimes V_{\phi}(I - H))B^{-1}_{\dagger} = (P \otimes  UK)\left(I_k \otimes K + P^{-1}(I_k - P) \otimes I_n\right)^{-1}(I_k \otimes U^{\T})$. As a result, 
\begin{equation*}
    \Cov_{DG}(\vect(\hat{B}_{NS}), \vect((X^{\T}X)^{-1}X^{\T}\hat{G}) \given X_1) = (A^{\T} \otimes (X^{\T}X)^{-1}X^{\T}U)M_1(A \otimes U^{\T}X(X^{\T}X)^{-1}),
\end{equation*}
where $M_1 = (P \otimes K)\left(I_k \otimes K + P^{-1}(I_k - P) \otimes I_n\right)^{-1}$. $M_1$ is diagonal with positive diagonal elements, so $\Cov_{DG}(\vect(\hat{B}_{NS}), \vect((X^{\T}X)^{-1}X^{\T}\hat{G}) \given X_1)$ is symmetric. We now turn to evaluation of $\Var_{DG}(\vect(\hat{G}) \given X_1)$. From \eqref{eqn:spatial_beta_estimate},
\begin{equation*}
    \Var_{DG}(\vect(\hat{G}) \given X_1) = (A^{\T} \otimes I_n)B^{-1}_{\dagger}(P \otimes (I_n - H)V_{\phi}(I_n - H) + (I_k - P) \otimes (I_n - H))B^{-1}_{\dagger} (A \otimes I_n). \\
\end{equation*}
Making the aforementioned substitutions for $V_{\phi}$, $I_n - H$, and $B^{-1}_{\dagger}$ yields 
\begin{equation*}
\begin{split}
    \Var_{DG}(\vect(\hat{G}) \given X_1) &= 
    (A^{\T} \otimes U)\left(I_k \otimes K + (P^{-1}(I_k - P)) \otimes I_n\right)^{-1}(P \otimes K^2  + (I_k - P) \otimes K) \\
    &\quad \cdot \left(I_k \otimes K + (P^{-1}(I_k - P)) \otimes I_n\right)^{-1} (A \otimes U^{\T}) 
\end{split}
\end{equation*}
The inner three terms of the product are diagonal matrices and commute. Thus,
\begin{equation*}
\begin{split}
    \Var_{DG}(\vect(\hat{G}) \given X_1) &= (A^{\T} \otimes U)\left(I_k \otimes K + P^{-1}(I_k - P) \otimes I_n\right)^{-2} \\
    &\quad\cdot(P \otimes K^2  + (I_k - P) \otimes K)(A \otimes U^{\T}),
\end{split}
\end{equation*}
Since $M_1 = \left(I_k \otimes K + (P^{-1}(I_k - P)) \otimes I_n\right)^{-2}(P \otimes K^2  + (I_k - P) \otimes K)$, then by \eqref{eqn:spatial_beta_estimate}, 
\begin{equation}\label{eqn:var_cond_spatial_estimator}
    \Var_{DG}(\vect(\tilde{B}_S) \given X_1) = \Var_{DG}(\vect(\hat{B}_{NS}) \given X_1) - (A^{\T} \otimes (X^{\T}X)^{-1}X^{\T}U)M_1(A \otimes U^{\T}X(X^{\T}X)^{-1}).
\end{equation}
Since $M_1$ is a positive definite matrix, $\Var_{DG}(\vect(\hat{B}_{NS}) \given X_1) - \Var_{DG}(\vect(\tilde{B}_S) \given X_1)$ is a positive definite matrix and $\Var_{DG}(\tilde{B}_{S,ij} \given X_1) < \Var_{DG}(\hat{B}_{NS,ij} \given X_1)$ for $i = 1, \ldots, p + 1$ and $j = 1, \ldots, k$. If $\Ep_{DG}[Z \given X] = XD$ as well, then $\Ep_{DG}[\tilde{B}_{S}] = \Ep_{DG}[\hat{B}_{NS}] = B + D$ and 
\begin{align*}
    \Var_{DG}(\tilde{B}_{S,ij}) &= \Ep_{X_1}[\Var_{DG}(\tilde{B}_{S,ij}\given X_1)] < \Ep_{X_1}[\Var_{DG}(\hat{B}_{NS,ij}\given X_1)] = \Var_{DG}(\hat{B}_{NS,ij}).
\end{align*}
By the bias-variance decomposition of mean squared error, the claim of \eqref{eqn:spatial_efficiency} follows. Similar calculations reveal that 
\begin{equation*}
    \begin{split}
        \Var_{DG}(\vect(\hat{B}_{NS}) \given X_1) &= A^{\T}(I_k - P)A \otimes (X^{\T}X)^{-1} \\ &\quad +(A^{\T} \otimes (X^{\T}X)^{-1}X^{\T}U)(P \otimes I_n)(A \otimes U^{\T}X(X^{\T}X)^{-1})
    \end{split}
\end{equation*}
Since $(P \otimes I_n) - M_1 = \left((I_k - P)^{-1} \otimes K + P^{-1} \otimes I_n\right)^{-1}$, then \eqref{eqn:postvar_cond_spatial_estimator} implies $\Var(\vect(\tilde{B}_S) \given X_1) = \Var(B_S \given Y, M = A, R = P, X_1)$.

\section{Metropolis within Gibbs Sampling for Multivariate BYM2 Model}\label{append:spatial_sampling_scheme}

\RestyleAlgo{ruled}
\begin{algorithm}[!ht]
\caption{Metropolis within Gibbs Sampling for MBYM2 Model}

\SetKwFunction{updateBeta}{updateBeta}
\SetKwFunction{updateGamma}{updateGamma}
\SetKwFunction{MHupdateM}{MHupdateM}
\SetKwFunction{MHupdateR}{MHupdateR}
\SetKwProg{Fn}{function}{:}{}

Given $Y, X, L_X = \texttt{chol}(X^{\T}X), I_n - H,  W_\phi^{-1}, v, \Sigma_0, \lambda_R, B^{(0)}, G^{(0)}, M^{(0)}, R^{(0)}, s_1, s_2, s_3$:\\
Initialize $M_1 \leftarrow X^{\T}Y; \quad M^{-1} \leftarrow (M^{(0)})^{-1}; \quad L \leftarrow \texttt{chol}(M^{(0)\T}M^{(0)})$.\\
Solve spectral decomposition $W_\phi^{-1} = Q \Lambda Q^{\T}$, $\Lambda = \mbox{diag}(\lambda_1, \ldots, \lambda_n)$.
Set $W_{\phi}^{-1/2} \leftarrow Q\Lambda^{1/2}Q^{\T}$.

\For{$t = 1, \ldots, T$} {
    Set $B^{(t)} \leftarrow$ \updateBeta{$G^{(t - 1)}, M^{(t - 1)}, P^{(t - 1)}$}.\\
    Set $G^{(t)} \leftarrow$ \updateGamma{$B^{(t)}, M^{(t - 1)}, M^{-1}, R^{(t - 1)}$}. \\
    Set $M^{(t)}, M^{-1}, L \leftarrow$ \MHupdateM{$B^{(t)}, G^{(t)}, M^{(t - 1)}, M^{-1}, L,  R^{(t - 1)}$} \\
    Set $R^{(t)} \leftarrow$ \MHupdateR{$B^{(t)}, G^{(t)}, M^{(t)}, M^{-1}, L, R^{(t - 1)}$}
}

\Fn{\updateBeta{$G, M, R$}}{
    \For{$j = 1, \ldots, k$} {
        \For{$i = 0, \ldots p$} {
            Sample $Z_{ij} \sim N(0, 1)$.
        }
    }
    Update $Z \leftarrow Z\mbox{diag}\left(\sqrt{1 - r_i}\right)M$.\\
    Set $B \leftarrow \texttt{forwardsolve}(L_{X}, M_1 - X^{\T}G)$; \quad $B \leftarrow \texttt{backsolve}(L_{X}^{\T}, B + Z)$.\\
    \KwRet $B$.
}

\Fn{\updateGamma{$B, M, M^{-1}, R$}}{
    Set $G \leftarrow Q^{\T}(Y - XB)M^{-1}$. \\
    \For{$j = 1, \ldots, k$} {
        \For{$i = 1, \ldots N$} {
            Sample $Z_{ij} \sim N(0, 1)$; \quad set $w_i \leftarrow \frac{1 - r_j}{r_j \lambda_i + 1 - r_j}$; \quad $z_i \leftarrow \sqrt{\frac{r_j(1 - r_j)}{r_j + (1 - r_j)\lambda_i}}$.
        }
        Update $G_{.j} \leftarrow G_{.j} \otimes w$; \quad $Z_{.j} \leftarrow Z_{.j} \otimes z$. ($G_{.j}$ denotes the $j$th column of $G$) \\ 
    }
    Update $G \leftarrow Q(G + Z)M$. \\
    \KwRet $G$.
}
\label{alg:sampling_loop}
\end{algorithm}

\RestyleAlgo{ruled}
\begin{algorithm}
\caption{Update steps for $M$ and $L$ in Metropolis within Gibbs sampling of MBYM2 Model}

\SetKwFunction{updateBeta}{updateBeta}
\SetKwFunction{updateGamma}{updateGamma}
\SetKwFunction{MHupdateM}{MHupdateM}
\SetKwFunction{MHupdateR}{MHupdateR}
\SetKwFunction{logLikGammaPrior}{logLikGammaPrior}
\SetKwFunction{logMPrior}{logMPrior}
\SetKwFunction{logRPrior}{logRPrior}
\SetKwFunction{Fbacksolve}{backsolve}
\SetKwFunction{Fforwardsolve}{forwardsolve}
\SetKwProg{Fn}{function}{:}{}

\Fn{\MHupdateM{$B, G, M, M^{-1}, L, R$}}{
    Initialize $M^{\star} \leftarrow 0_{k\times k}$; \quad $\alpha_{MH} \leftarrow 0$; \quad sample $u \sim \text{Unif}(0, 1)$. \\
    \For{$j = 1, \ldots, k$} {
        \For{$i = j, \ldots, k$} {
            \lIf{$i = j$}{sample $\log(M^{\star}_{ii}) \sim N(\log(M_{ii}), s_1^2)$.}
            \lElse{sample $M^{\star}_{ij} \sim N(M_{ij}, s_2^2)$.}
        }
    }
    Set $(M^{\star})^{-1} \leftarrow \texttt{forwardsolve}(M^{\star}, I_k)$.\\
    Update $\alpha_{MH} \leftarrow \alpha_{MH} + \logLikGammaPrior(B, G, M^{\star}, (M^{\star})^{-1} , R) + \logMPrior(M^\star)$.\\
    Update $\alpha_{MH} \leftarrow \alpha_{MH} - \logLikGammaPrior(B, G, M, M^{-1}, R) - \logMPrior(M)$. \\
    \DontPrintSemicolon
    \lIf{$\log(u) \leq \alpha_{MH}$}{\KwRet $M^\star, (M^{\star})^{-1} $.}
    \lElse{\KwRet $M, M^{-1}$.}
}

\Fn{\MHupdateR{$B, G, M, M^{-1}, L, R$}}{
    Initialize $\alpha_{MH} \leftarrow 0$; \quad sample $u \sim \text{Unif}(0, 1)$.  \\
    \For{$i = 1, \ldots k$}{
        Sample $\tau \sim N\left(\log\left(\frac{r_i}{1 - r_i}\right), s_3^2\right)$; \quad $r^{\star}_i \leftarrow \frac{\exp(\tau)}{1 + \exp(\tau)}$. \\
        Update $\alpha_{MH} \leftarrow \alpha_{MH} + \log\left(r_i(1 - r_i)\right) - \log\left(r_i(1 - r_i)\right)$.
    }
    Update $\alpha_{MH} \leftarrow \alpha_{MH} + \logLikGammaPrior(B, G, M, M^{-1}, R_{\star}) + \logRPrior(R_\star)$.\\
    Update $\alpha_{MH} \leftarrow \alpha_{MH} - \logLikGammaPrior(B, G, M, M^{-1}, R) - \logRPrior(R)$. \\
    \DontPrintSemicolon
    \lIf{$\log(u) \leq \alpha_{MH}$}{\KwRet $R_\star$.}
    \lElse{\KwRet $R$.}
}

\label{alg:update_steps}
\end{algorithm}

\RestyleAlgo{ruled}
\begin{algorithm}[!ht]
\caption{Log-likelihood and prior densities for Metropolis acceptance ratios}

\SetKwFunction{Fbacksolve}{backsolve}
\SetKwFunction{Fforwardsolve}{forwardsolve}
\SetKwProg{Fn}{function}{:}{}

\Fn{\logLikGammaPrior{$B, G, M, M^{-1}, R$}}{
    Initialize $d \leftarrow 0$; \quad $\Delta \leftarrow Y - XB - G$; \quad $\theta \leftarrow W^{-1/2}_{\phi} G M^{-1}$. \\
    \For{$j = 1, \ldots, k$} {
        \For{$i = 1, \ldots, N$} {
            Update $d \leftarrow d - \frac{1}{2}\left(\frac{\Delta_{ij}^2}{1 - r_j} + \frac{\theta_{ij}^2}{r_j}\right)$.\\
        }
        Update $d \leftarrow d - \frac{N}{2}\log\left(r_j(1 - r_j)\right) - 2N \log(M_{jj})$.\\
    }
    \KwRet $d$.
}

\Fn{\logMPrior{$L$}} {
    Initialize $d \leftarrow 0$. \\
    Set $V \leftarrow \texttt{forwardsolve}(M, \Sigma_0)$; \quad update $V \leftarrow \texttt{backsolve}(M^{\T}, V)$. \\
    \For{$i = 1, \ldots, k$} {
        Update $d \leftarrow d - (v + i) \log(M_{ii}) - \frac{v}{2} V_{ii}$.
    }
    \KwRet $d$.
}

\Fn{\logRPrior{R}}{
    Initialize $d \leftarrow -\left(\sum_{i = 1}^{k} r_i\right) \cdot \left(N - \sum_{j = 1}^N \lambda_{j}^{-1} \right) $. \\
    \For{$i = 1, \ldots, k$} {
        \For{$j = 1, \ldots N$} {
            Update $d \leftarrow d - \log\left(1 - r_i + r_i / \lambda_j\right)$. \\
        }
    }
    \KwRet $-\lambda_R \sqrt{d}$.
}

\label{alg:density_computations}
\end{algorithm}

Algorithm~\ref{alg:sampling_loop} details a sampling scheme that applies the Gibbs update steps for $\{B, G\}$ and Metropolis update steps for $\{M, R\}$ provided in Algorithm~\ref{alg:update_steps} for the multivariate BYM2 model in \eqref{eqn:spatial_analysis_model} when the prior is given by \eqref{eqn:full_spatial_prior}. Algorithm~\ref{alg:sampling_loop} takes as inputs the data $\{X, Y\}$, prior parameters $\{W_{\phi}^{-1}, v, \Sigma_0, \lambda_R\}$, initial values $\{B^{(0)}, G^{(0)}, M^{(0)}, R^{(0)}\}$, and proposal parameters $\{s_1, s_2, s_3\}$. Here, $\texttt{chol}$ returns the lower triangular Cholesky factor of a positive definite matrix, $\texttt{updateBeta}$ returns $B_S$ from $\vect(B_S) \given Y, G, A, P, X_1 \sim N_{pk}\left(\vect(\mu_{B}), V_B \right)$, where $\mu_{B} = (X^{\T}X)^{-1}X^{\T}(Y - G)$ and $V_{B} = M^{\T}(I_k - R)M \otimes (X^{\T}X)^{-1}$, while $\texttt{updateGamma}$ returns $G$ from $\vect(G) \given Y, X_1, B_S, A, P \sim N_{nk}(F_{G}f_{G}, F_{G})$, where 
\begin{equation*}
\begin{split}
    F_{G} &= (M^{\T} \otimes I_n)\left((I_k - R)^{-1} \otimes I_n + R^{-1} \otimes W_{\phi}^{-1}\right)^{-1}(M \otimes I_n), \\ f_{G} &= (M^{-1}(I_k - R)^{-1} M^{-\T} \otimes I_n) \vect(Y - XB_S).
\end{split}
\end{equation*}
$\texttt{MHupdateM}$ updates the lower triangular matrix $M$ via a Metropolis random walk by proposing the lower triangular $k\times k$ $M^{\star}$, constructed by sampling diagonal elements $\log(M^{\star}_{ii}) \sim N(\log(M_{ii}), s^2_1)$ and off-diagonal elements $M^{\star}_{ij} \sim N(M_{ij}, s^2_2)$. 
If $M^{\star}$ is accepted, $(M^{\star})^{-1}$ is also returned for intermediate calculations. Finally, $\texttt{MHupdateR}$ performs a Metropolis update of $R$ by proposing $R^{\star} = \mbox{diag}(r^{\star}_1, \ldots, r^{\star}_k)$ where $\log\left(\frac{r^{\star}_i}{1 - r^{\star}_i}\right) \sim N\left(\log\left(\frac{r^{\star}_i}{1 - r^{\star}_i}\right), s^2_3\right)$. To compute the Metropolis acceptance ratios, $\texttt{MHupdateM}$ and $\texttt{MHupdateR}$ 
call the functions in Algorithm \ref{alg:density_computations} to compute posterior densities up to a constant. \texttt{logLikGammaPrior} computes $\log \pi(Y \given B_S, G, M, R, X_1) + \log\pi(G \given M, R) + C_1$, $\texttt{logMPrior}$ evaluates $\log\pi(M) + C_2$, and $\texttt{logRPrior}$ returns $\log\pi(R) + C_3$. Here, 
\begin{equation*}
\begin{split}
    \pi(Y\given B_S, G, M, R, X_1) &= N_{nk}\left(\vect(Y) \given \vect(XB_S + G), M^{\T}(I - R)M \otimes I_n \right), \\
    \pi(G \given M, R) &= N_{nk}\left(\vect(G) \given 0_{nk}, M^{\T}RM \otimes W_{\phi} \right), \\
    \pi(M) &= \mbox{IW}(MM^{\T} \given v, v\Sigma_0) \times \prod_{i = 1}^{k} M_{ii}^{k - i + 1},\\
    \pi(R) &= \mbox{PC}(R \given \lambda_R),
\end{split}
\end{equation*}
where $\mbox{PC}(R \given \lambda_R)$ is a penalized complexity prior discussed in Section~\ref{sec:dg_analysis_models} and $\{C_1, C_2, C_3\}$ is a set of constants that does not depend on model parameters.


\section{Health Data Setup and Non-Spatial Analysis}\label{append:RDA_setup}

In Section~\ref{sec:application}, we analyze prevalence data on obesity and diabetes alongside diabetes-related cancer mortality rate estimates in US counties. Estimates of diabetes and obesity age-adjusted prevalence rates among adults aged 20 years or more for 3,142 US counties in 2015 were extracted from the US Diabetes Surveillance System, a database maintained by the \cite{centersfordiseasecontrolandpreventionUSDiabetesSurveillance2015} (CDC). We define diabetes-related cancer mortality with an underlying cause of death related to liver, pancreatic, endometrial, colorectal, breast, or bladder cancer \citep{giovannucciDiabetesCancer2010}. We obtained age-adjusted mortality rates aggregated from 2010 to 2020 for 3,008 counties based on US resident death certificates classified with the corresponding ICD-10-CM (International Classification of Diseases, Tenth Revision, Clinical Modification) code from the 1999-2020 Underlying Cause of Death WONDER database, made publicly available by the \cite{centersfordiseasecontrolandpreventionnationalcenterforhealthstatisticsMortality19992020CDC2021}.

We account for the economic context of counties through the median household income, poverty rate, and unemployment rate in 2015. The healthcare context is represented by the percentage of the population under 65 years of age without health insurance, the number of primary care physicians per 100,000 people, and the number of outpatient visits per 100,000 people in 2015. Economic and healthcare variables were extracted from the US Department of Health and Human Services 2019-2020 County Level Area Health Resources Files (AHRF), which includes historical data from previous years. 

The natural and built environment encapsulates both the food environment and the physical environment, otherwise known as the recreational context. We quantify the food environment by the percentage of the population living in a food desert and the percentage of the population receiving food stamp or Supplemental Nutrition Assistance Program (SNAP) benefits, both estimated for 2015. A person is defined as living in a food desert or low-access area if they live more than one mile from a supermarket or large grocery store in an urban area or more than ten miles in a rural area. Estimates of low access rates were extracted from the Food Environment Atlas, made publicly available by \cite{economicresearchserviceersu.s.departmentofagricultureusdaFoodEnvironmentAtlas2020} of the US Department of Agriculture (USDA). The percentages of food stamp or SNAP recipients were drawn from the 2019-2020 AHRF. The recreational context was captured by the age-adjusted percentage of adults aged 18 years or older in the US who are physically inactive in 2015, and the number of recreation and fitness facilities per 1,000 people in 2016. A person is defined as physically inactive if they obtain less than 10 minutes per week of moderate or vigorous activity in each category of physical activity of work, leisure time, and transportation. Data on physical inactivity were drawn from the CDC's US Diabetes Surveillance System, while data on recreation and fitness facilities was taken from the USDA's Food Environment Atlas.

The population structure was captured by six variables: the percentage of the population that was non-Hispanic black, the percentage of the population that was Hispanic, the percentage of the population at least 65 years old, the percentage of the population at most 18 years old, the percentage of the population living in urban areas, and the percentage of adults at least 25 years old that hold a high school diploma. Population structure data was extracted from the 2019-2020 AHRF, where all covariates except education were originally sourced from the 2010 Census of the US Census Bureau. The percentage of adults at least 25 years old that hold a high school diploma was sourced from the US Census Bureau's 2011-2015 American Community Survey's five-year estimates.

We first conduct a non-spatial analysis using the non-spatial model in \eqref{eqn:nonspatial_analysis_model}. Following the prior recommendations of \cite{jinOrderFreeCoRegionalizedAreal2007}, we set $v = 3$ and $\Sigma_0$ as a diagonal matrix where the $j$th diagonal element is $||(I_n - H)Y_j||^2 / (n - p - 1)$. We obtain 10,000 exact posterior samples of $B_{NS}$ and $\Sigma_{NS}$ by simulating $\Sigma_{NS} \sim \mbox{IW}(v_\star, \Sigma_\star)$ and then drawing one instance of $B_{NS} \sim \mbox{N}_{pk}\left(\vect\left(\hat{B}_{NS}\right), \Sigma_{NS} \otimes (X^{\T}X)^{-1}\right)$ for each sampled value of $\Sigma_{NS}$. 

\begin{table}[ht]
\centering
\resizebox{\columnwidth}{!}{\begin{tabular}{llll}
  \hline
Variable & Diabetes & Obesity & Cancer Mortality \\ 
  \hline
Intercept & 5.9e-06 (-0.022, 0.024) & -1.9e-04 (-0.024, 0.025) & -3.9e-05 (-3e-02, 3e-02) \\ 
  Poverty Rate & -0.043 (-0.1, 0.017) & -6e-02 (-0.12, 7.8e-03) & -0.042 (-0.12, 0.037) \\ 
  Median Income & -0.049 (-0.099, 3.3e-04) & -0.19 (-0.24, -0.13) & -9e-02 (-0.16, -0.026) \\ 
  Unemployment & 0.1 (0.068, 0.14) & -0.014 (-0.049, 0.024) & -0.11 (-0.15, -0.064) \\ 
  Uninsured Rate & 0.027 (-7.7e-03, 6e-02) & -0.29 (-0.32, -0.25) & -0.058 (-0.1, -0.014) \\ 
  PCP Density & -0.016 (-0.046, 0.015) & -0.016 (-0.047, 0.017) & -0.069 (-0.11, -0.028) \\ 
  Outpatient Visits & 0.023 (-3.2e-03, 0.049) & 0.032 (4.2e-03, 6e-02) & 0.042 (9.7e-03, 0.079) \\ 
  Low Access & 0.015 (-0.011, 0.038) & 0.046 (0.021, 0.073) & 0.047 (0.015, 8e-02) \\ 
  SNAP Assistance & 0.17 (0.12, 0.22) & 2e-03 (-0.054, 0.053) & 0.31 (0.25, 0.38) \\ 
  Physical Inactivity & 0.51 (0.48, 0.53) & 0.6 (0.58, 0.63) & 0.089 (0.054, 0.12) \\ 
  Recreation Facilities & -0.057 (-0.084, -0.029) & -2e-02 (-0.049, 8.4e-03) & -0.028 (-0.064, 7.5e-03) \\ 
  Percent NH-Black & 0.19 (0.16, 0.22) & 0.061 (0.027, 0.092) & 0.14 (0.1, 0.19) \\ 
  Percent Hispanic & -0.084 (-0.12, -0.051) & -0.1 (-0.14, -0.068) & -0.16 (-0.2, -0.11) \\ 
  Percent $\geq$ 65 years & -0.11 (-0.15, -0.074) & 5.5e-03 (-0.032, 0.044) & 0.099 (0.052, 0.15) \\ 
  Percent $\leq$ 18 years & 8.5e-03 (-0.023, 0.043) & 0.16 (0.12, 0.19) & 0.072 (0.032, 0.12) \\ 
  Urban Percent & 0.11 (0.073, 0.14) & 0.035 (-2.1e-03, 0.071) & 0.092 (0.047, 0.14) \\ 
  HS Diploma Rate & 8.1e-03 (-0.034, 0.053) & 0.028 (-2e-02, 0.072) & -0.21 (-0.27, -0.15) \\ 
   \hline
\end{tabular}}
\caption{Non-spatial model regression coefficient posterior means and 95\% highest posterior density credible regions.} 
\label{tab:nonspatial_model_estimates}
\end{table}

Table~\ref{tab:nonspatial_model_estimates} lists the standardized coefficient estimates and 95\% credible intervals for the non-spatial analysis model, showing differences in the set of significant predictors when compared to the spatial analysis in Section~\ref{sec:application}. For the diabetes prevalence coefficients, outpatient visit rate and percent of residents 18 years old or younger are significant in the spatial model but not in the non-spatial model, whereas percentage of Hispanic residents is significant in the non-spatial model but not in the non-spatial model. For the obesity prevalence coefficients, primary care physician density, percent of residents at least 65 years old, and percent of residents in urban areas are significant in the spatial model but not in the non-spatial model, while outpatient visit rate and percent of residents living in food deserts are significant in the non-spatial model but not in the spatial model. For diabetes-related cancer mortality, the set of significant predictors is identical between the spatial and non-spatial analysis models.

\bibliography{spatial_confounding}
\bibliographystyle{abbrvnat}